\documentclass[a4paper]{article}
\usepackage{graphicx}
\usepackage{subfigure}
\usepackage{amsmath}
\usepackage{amssymb}
\usepackage[numbers,sort]{natbib}
\usepackage{url}
\usepackage{multirow}
\usepackage{array}
\usepackage{booktabs}
\usepackage{tabularx}
\usepackage{lscape}
\usepackage{verbatim}
\usepackage[table]{xcolor}
\newcommand{\newrow}[9]{
\hline
\rowcolor[gray]{#9}\footnotesize #1 & \footnotesize #2 & \footnotesize #3 & \footnotesize #4 & \footnotesize #5 & \footnotesize #6\\
\hline
\rowcolor[gray]{#9}\multicolumn{6}{|p{.9625\hsize}|}{\footnotesize #7}\\
\hline
\rowcolor[gray]{#9}\multicolumn{6}{|p{.9625\hsize}|}{\footnotesize #8}\\\hline
\noalign{\smallskip}
}

\def\beq{\begin{equation}}   \def\eeq{\end{equation}}
\def\bea{\begin{eqnarray}}   \def\eea{\end{eqnarray}}

\newcommand{\state}[1]{| 1\rangle}

\newcommand{\gsim}{\lower.7ex\hbox{$ \;\stackrel{\textstyle>}{\sim}\;$}}
\newcommand{\lsim}{\lower.7ex\hbox{$ \;\stackrel{\textstyle<}{\sim}\;$}}

\def\c2{CLEO~II.V}

\def\d0d0{ D^0\bar{D}^0 }
\def\p0p0{ P^0\bar{P}^0 }

\def\qp2{ \Bigl| \frac{q}{p} \Bigr|^2 }
\def\pq2{ \Bigl| \frac{p}{q} \Bigr|^2 }

\def\ps2s{  \psi(2S) }
\def\q2{ $q^2$ }
\def\cm2s1{ $\,{\rm cm}^{-2} {\rm s}^{-1}$}
\def\d0{D_2^{*0}}
\def\d+{D_2^{*+}}

\newcommand{\nwchem}{{\sc NWChem }}
\newcommand{\pymol}{{\sc PyMol }}

\newcommand{\Header}{
  \begin{tabular}{rl}
  \hspace{-.4cm}
      &
    \renewcommand{\arraystretch}{0.5}
    \renewcommand{\arraystretch}{1}
  \end{tabular}
  \vskip 1cm
  \begin{flushright}
  \renewcommand{\arraystretch}{0.5}
    \begin{tabular}{r}
      {\underline{ INFN-14-13/LNF}}\\    
     {\underline{ CERN-OPEN-2015-004}}\\
    \end{tabular}
  \end{flushright}
  \renewcommand{\arraystretch}{1}
  \vskip 1 cm
  }
\begin{document}
\begin{titlepage}

\title{
  \Header{
    \LARGE 
     \bf  PROPERTIES OF POTENTIAL ECO-FRIENDLY GAS REPLACEMENTS FOR PARTICLE
     DETECTORS IN HIGH-ENERGY PHYSICS}
      }
\author{
L.~Benussi$^a$,
S.~Bianco$^a$,
D.~Piccolo$^a$,
G.~Saviano$^b$
\\
S.~Colafranceschi$^c$,
J.~Kj$\o$lbro$^c$,
A.~Sharma$^c$
\\
D.~Yang$^d$,
G.~Chen$^d$,
Y.~Ban$^d$,
Q.~Li$^d$
} 

\maketitle
\baselineskip=1pt
\begin{abstract}
\indent 
   Modern gas detectors for detection of particles require F-based
   gases for optimal performance. Recent regulations demand the use of
    environmentally unfriendly F-based gases to be limited or banned.
    This review studies properties of potential eco-friendly gas
    candidate replacements.
\end{abstract}
\vspace*{\stretch{2}}
\begin{flushleft}
\end{flushleft}
\begin{center}
\emph{Submitted to Journal of Instrumentation}
\end{center}
\vskip 0.5cm
\begin{flushleft}
\vskip 0.5cm
\begin{tabular}{l l}
  \hline\\
  $ ^{a\,\,\,}$& \footnotesize{Laboratori Nazionali di Frascati dell'INFN, Italy} \\
  $ ^{b}$ & \footnotesize{Laboratori Nazionali di Frascati dell'INFN and Facolta' di Ingegneria Universit\`a di Roma La Sapienza, Italy} \\
  $ ^{c}$ & \footnotesize{CERN, Geneva, Switzerland} \\
  $ ^{d}$ & \footnotesize{Peking University, Beijing, China} \\
\end{tabular}
\end{flushleft}
\end{titlepage}
\pagestyle{plain}
\setcounter{page}2
\baselineskip=17pt
\section{Introduction}
Many currently used refrigerant gases have a great impact on the environment since they either contribute largely to the greenhouse gas effect, or because they tear the ozone layer, or both. In an attempt to protect the environment, regulations preventing the production and use of certain refrigerant gases have been implemented \cite{EPA}.
\par
Gas detectors are widespread for detection, tracking and triggering of charged particles such as muons in Nuclear and High Energy Physics (HEP). They are characterized by simple and reliable use, but utmost care must be taken to issues such as  properties of gas interaction with materials, gas purification, gas mixture contaminants, etc. 
\cite{Abbrescia:2007mu}-
\cite{Benussi:2010yx}.
\par
 A large part  of  gas muon detectors used in HEP  operates with mixtures containing the regulated refrigerants as quenching medium in applications where excellent time resolution and avalanche operation are necessary. Therefore,  actions towards finding new mixtures must be undertaken. Gas Electron Multiplier (GEM)  \cite{Sauli:1999ya}
   detectors  operate in experiments such as CMS (Compact Muon Solenoid) at the LHC (Large Hadron Collider)  with an argon/CO$_2$  mixture \cite{Sharma:2012zzb}. However,  for high time resolution applications  an argon/CO$_2$/CF$_4$ mixture 
   is used \cite{Alfonsi:2006ny},
   where CF$_4$ has a Global-Warming Potential (GWP) of 7390 \cite{GWPipcc}. 
   Resistive Plate Counters (RPC)  \cite{Santonico:1981sc}
   currently operate with a F-based R134a/Isobutane/SF$_6$ gas mixture, with typical GWP of 1430. Investigations into new gas mixtures has to be performed in order to keep the mixture properties while complying with the regulations. A few industrial refrigerant industrial replacements were proposed \cite{Abbrescia:2013} as alternatives to R134a. A study of transport properties of currently used gas mixtures in HEP, and evaluation of transport properties of freon-less gas mixtures,  was recently published \cite{Assran:2011Thesis}\cite{Assran:2011ug}.
\par
The aim of this paper is to discuss some of the important properties of gases for particle gas detectors, to list and summarize basic properties of eco-friendly refrigerants from the literature, to discuss their properties for materials compatibility and safe use, and make a prediction on selected parameters crucial for the performance of gas detectors considered. While this study is aimed to GEM and RPC detectors, its findings can be considered for selection of ecogas replacement for other gas detectors.
\section{Gas properties}
For a gas mixture to be appropriate in a gas detector, first of all it has to comply with the regulations. Furthermore, its properties must also be appropriate for the specific type of detectors. For example, a gas that is suitable for the RPC detectors may not be fully optimized for the GEM detectors.  To better find the appropriate gas for a detector, an understanding of the influence of different parameters is required. This section aims to clarify the most essential parameters for gases.
\subsection{Global-warming potential and Ozone depletion potential}
%
%
In order to estimate the impact of a refrigerant on the environment, the effects have to be quantified. Two important effects are the contribution to the greenhouse effect and the depletion of the ozone layer. The first mentioned effect is measured in Global-Warming Potential (GWP), and is normalized to the effect of CO$_2$ (GWP = 1), while the effect on the ozone layer is measured in Ozone Depletion Potential (ODP), normalized to the effect of CCl$_3$F (ODP = 1). The effects of selected refrigerant candidates are listed in table \ref{table:summary}.

\subsection{Stopping power}
When a particle passes through a medium, energy is transferred from the particle to the surroundings. The energy lost is typically defined as  the stopping power expressed as $\dfrac{1}{\rho}\left<\dfrac{dE}{dx}\right>$, where $\rho$ denotes the density of the medium, E denotes energy, and x is length. This expression is itself dependent on the momentum $p$ of the particle. In the next section it will be shown the particle will experience a minimum energy loss at around $p/Mc\approx 3$, where $M$ is the particle mass, and $c$ is the speed of light. By knowing the average minimum energy loss, we also know the minimum energy the particle loses by passing through a detector. The minimum mean ionization energies for the refrigerants under consideration are summarized in table \ref{table:calculated_quantities}.
\subsection{Radiation length}
The radiation length $X_0$ is a characteristic length of a medium. It describes both the mean distance required for a high energy electron to lose all but $e^{-1}$ of its energy due to bremsstrahlung, and ${}^7\!/_{9}$ of the mean free path of a $e^+e^-$ pair produced by a high-energy photon
\cite{Tsai:1973py}. These quantities are also estimated and summarized in table \ref{table:calculated_quantities}.
\subsection{Ion pair production}
When an incoming particle passes through a medium, it will eventually interact with the medium and transfer some of its energy to ionize atoms. In this process, a pair consisting of an ionized atom and a free electron is produced. The number of ionizations produced by an incoming particle per unit length is denoted by $N_P$, in units of cm$^{-1}$. Each produced ion pair will have an initial kinetic energy and can itself produce an ion pair, called secondary ion pair production. The sum of the primary and secondary ion pairs production per unit length is denoted $N_T$, and will be mainly depending on the material and the incoming particle energy and mass. This parameter is relevant in particle gas detectors as it determines both the number  and the size of avalanches produced by a single incoming particle when the gas is under an amplifying electric field.
\subsection{Drift velocity}
In the absence of electric field, electrons move randomly in all direction having an average thermal energy $3/2 KT$. In presence of an electric field, the electrons start to drift along the field direction with mean drift velocity $v_d$ (the average distance covered by the drift electron per unit time). The drift velocity depends on the pressure, temperature and could be affected by the presence of pollutants like water or oxygen. Electric field lines are, in addition, deformed by any magnetic field applied to the gas volume. Finally, electronegative pollutants deplete the gas of electrons. Computer codes such as Magboltz \cite{Magboltz} computes the drift velocity using montecarlo techniques, by tracing electrons at the microscopic level through numerous collisions with gas molecules.
\subsection{Diffusion}
Electrons and ions in a gas are subject only to an electric field and move on average along the electric field, but individual electrons deviate from the average due to scattering on the atoms of the gas. Scattering leads to variations in velocity, called longitudinal diffusion, and to lateral displacements, called transverse diffusion. The scattering process in each direction can, to a good approximation, be considered gaussian on a microscopic scale. In cold gases like carbon-dioxide for example, the diffusion is small, while the drift velocity is low and unsaturated for values of electric fields common in gas detectors; this implies a non linear space time relation. Warm gases, like argon for instance, have a higher diffusion; when mixed with polyatomic/organic gases having vibrational thresholds between 0.1 and 0.5 eV, diffusion is reduced in most cases, while the drift velocity is increased.
\subsection{Townsend coefficients}
The average distance an electron travels between ionizing collisions is called mean free path and its inverse is the number of ionizing collision per centimeter $\alpha$ (the first Townsend coefficient). This parameter determines the gas gain of the gas. If $n_0$ is the number of primary electron without amplification in uniform electric field, and $n$ is the number of electrons after distance $x$ under avalanche condition, then $n$ is given by $n = n_0 e^{\alpha x}$ and the gas gain $G$ is given by 
$G \equiv n_0/n = e^{\alpha x}$. The first Townsend coefficient depends on the nature of the gas, the electric field and pressure. To take into account the augmented emission of electrons by the cathode caused by impact 
of positive ions, it is customary to introduce $\eta$,  Townsend's second ionization coefficient or 
attachment parameter,
{\it i.e.}, the average number of electrons released from a surface by an incident positive ion, according to the formula
\begin{equation}
G \equiv \frac{e^{\alpha x}}{1-\eta(e^{\alpha x}-1)}
\end{equation}
\subsection{Lorentz angle}
Because of the deflection effect exerted by a magnetic field perpendicular to the electric field and the motion of the electron, the electron moves in a helical trajectory resulting in a lowered drift velocity and transverse dispersion. Thus the arrival time of electrons at the anode changes and the spread in the drift time increases. The angle which the drifting electron swarm makes with the electric field is defined as the Lorentz angle of the particular gas or gas mixture under consideration. The Lorentz angle depends on both the electric field and the magnetic field. It is normally large at small electric fields but falls to smaller values for larger electric fields and is approximately linear with increasing magnetic field.
\subsection{Hazards and flammability}
Many refrigerants may constitute danger for the user and its environment. The greatest dangers involved are the flammability and toxicity. We have used two standards in categorizing the refrigerants in table \ref{Table:1} and \ref{Table:2}. The Ashrae standard \cite{Ashrae} gives each refrigerant a number denoting flammability from 1 (not flammable) to 3 (highly flamable), as well as a letter A (non-toxic) or B (Toxic). The Health Material Hazardous Material Information System (HMIS), rates Health/ Flammability/ and Physical hazards from 0 (low) to 4 (high).
\subsection{Compatibility with materials}
Some refrigerants are incompatible with certain materials, and can either react violently, or have long term effect. Some refrigerants may even produce toxic decomposition and/or polymerization. Known incompatibilities and toxic byproducts are summarized in Table \ref{Table:1} and \ref{Table:2}.
\subsection{Aging and  longevity}
Aging is defined (following \cite{Va'vra:2002yd}) the general deterioration of the detectors during their operation. The aging phenomenon is very complex and depends on several parameters. The commonly used variables include the cross-sections, electron/photon energies, electrostatic forces, dipole moments, chemical reactivity of atoms and molecules, etc. For a comprehensive (although non recent) collection see \cite{Sauli:2001ad}\cite{Schmidt:2001ae}\cite{Hohlmann:2001af}. A more recent review of ageing effects in GEM detectors can be found in \cite{Merlin:2014Thesis}.
\begin{table}
\centering
\footnotesize
\def\arraystretch{1.3}
\begin{tabularx}{1\hsize}{p{0.26\hsize} X X X X X}
Molecular name & Chemical formula & CAS & Refrigerant identifier & GWP & ODP\\
\hline
Chloropentafluoroethane &$C_2ClF_5$&76-15-3&R115&7370&0.44\\
Hexafluoroethane &$C_2F_6$&76-16-4&R116& - & - \\
2,2-Dicloro-1,1,1-trifluoroethane &$C_2HCl_2F_3$&306-83-2&R123&120&0\\
1-Chloro-3,3,3-Trifluoropropene &$C_3H_2ClF_3$&2730-43-0&R1233zd&4.7-7&0\\
2,3,3,3-Tetrafluoropropene &$C_3H_2F_4$&754-12-1&R1234yf&4&0\\
1,3,3,3 Tetrafluoropropene &$C_3H_2F_4$&29118-24-9&R1234ze&6&0\\
Trifluoroiodomethane &$CF_3I$&2314-97-8&R13I1& 0&0\\
1,1,1,2-Tetrafluoroethane &$CH_2FCF_3$&811-97-2&R134a&1430&0\\
Tetrafluoromethane &$CF_4$&75-73-0&R14& 7390 & 0 \\
1,1,1-trifluoroethane &$CH_3CF_3$&420-46-2&R143a& 4300& - \\
1,1-Difluoroethane&$C_2H_4F_2$&75-37-6&R152a&124&0\\
Octafluoropropane &$C_3F_8$&76-19-7&R218& - & - \\
Propane &$C_3H_8$&74-98-6&R290&3&0\\
Difluoromethane &$CH_2F_2$&75-10-5&R32&650&0\\
Isobutane &$C_4H_{10}$&75-28-5&R600a&3&0\\
Sulfur Hexafluoride &$SF_6$& 2551-62-4&R7146&~23000&0.04\\
Carbon Dioxide &$CO_2$&124-38-9&R744&1&0\\
Octafluorocyclobutane &$C_4F_8$&115-25-3&RC318& - & - \\
Pentafluoroethane & $HF_2CF_3$ & 354-33-6 & R125 & 3400 & 0  \\
Trifluoromethane  & $CHF_3$ & 75-46-7 & R23 & 0 & 0    \\
R409 : & $CHClF_2$ & 75-45-6 2837-89-0 75-68-3 & R22 (60\%), R142b (25\%), R124 (15\%)  & 1700 - 620 & 0.5 / 0.065 / 0.02 \\
R407c :  & $CH_2F_2$, $CF_3CHF_2$, $CH2FCF_3$ & 75-10-5, 354-33-6, 811-97-2 & R32 (21-25\%), R125 (23-27\%), R134a (50-54\%) & 650 3400 1430  & 0 0 0 \\
\end{tabularx}
\caption{Summary of various refrigerant candidates.}
\label{table:summary}
\end{table}
\begin{table}
\centering
\begin{tabular}{l | c c c c}
Name & I & $-\left<\dfrac{dE}{dx}\right>_{min} $ & $X_0 $ & $N_P$\\ \tiny  & & &  \\ \normalfont 
& $\left[\text{eV}\right]$ & $\left[\text{MeV} \dfrac{\text{g}}{\text{cm}^2}\right]$ & $\left[\dfrac{\text{g}}{\text{cm}^2}\right]$ &$\left[\text{cm}^{-1}\right]$\\
\hline\hline
R32 & 89.3602 & 1.80973 & 35.4581 & 49.2\\
R7146 & 127.401 & 1.67833 & 28.6027 & 92.0\\
R600a & 47.848 & 2.24057 & 45.2260  & 81.0\\
R1234yf & 91.9674 & 1.7734 & 35.8204  & 89.5\\
R152a & 78.1889 & 1.88706 & 37.0969 & 67.1\\
R1234ze & 91.9674 & 1.7734 & 35.8204  & 89.5\\
R115 & 116.695 & 1.69178 & 29.2197 & 98.4\\
R1233zd & 106.689 & 1.73915 & 29.7636  & 105\\
R290 & 47.0151 & 2.26184 & 45.3725 & 65.2\\
R13\textbar 1 & 271.737 & 1.42486 & 11.5399  & 272\\
R134a & 95.0294 & 1.76439 & 35.1542 & 81.6\\
R14 & 107.127 & 1.69909 & 33.9905  & 63.6\\
R123 & 125.275 & 1.69722 & 25.5416 & 98.4\\
R143a & 87.8152 & 1.8126 & 35.8928  & 74.8\\
R744 & 88.7429 & 1.81124 & 36.1954 & 37.2\\
R23 & 99.9508 & 1.7402 & 34.5214 & 56.9\\
R116 & 105.075 & 1.70566 & 34.2947 & 93.3\\
RC318 & 101.578 & 1.71721& 34.8435 & 123\\
R218 & 104.13 & 1.70873 & 34.439 & 117
\end{tabular}
\caption{Minimum ionization, radiation length and number of primary ion pair creation for the considered refrigerants, as well as the approximated mean ionization energy used.}
\label{table:calculated_quantities}
\end{table}
\begin{table}
\centering

\begin{tabularx}{1\textwidth}{|X|X|X|X|X|X|}


\newrow{\textbf{Refrigerant}}{\textbf{Molecular weight}}{\textbf{Density g/L}}{\textbf{Boiling point degC}}{\textbf{HMIS}}{\textbf{Ashrae Safety Group}}{\textbf{Material incompatibility}}{\textbf{Hazardous decomposition products and polymerization}}{0.9}

\newrow{R115}{154.4}{6.623}{-39.1}{1/0/2}{A1}
{Material is stable. However, avoid open flames and high temperatures. Incompatible with alkali or alkaline earth metals-powdered Al, Zn, Be, etc.}
{Decomposition product are hazardous."FREON" 115 Fluorocarbon can be decomposed by high temperatures (open flames, glowing metal surfaces, etc.)forming hydrochloric and hydrofluoric acids, and possibly carbonyl halides. Thermal decomposition can yield toxic fumes of fluorides such as Hydrogen Fluoride,  Hydrogen Chlo- ride, Carbon Monoxide and Chlorine.}{1}

\newrow{R116}{138.01}{5.734 }{-79}{ 1/0/0 }{A1}
{May  react violently with alkaline-earth and alkali metals Thermal decomposition yields toxic products which can be corrosive in the presence of moisture. If involved in a fire the following toxic  and/or corrosive fumes may be produced by thermal decomposition: Carbonyl fluoride, Hydrogen fluoride, Carbon monoxide }
{Thermal decomposition products: halogenated compounds, oxides of carbon  }{1}

\newrow{R1233zd}{130.5}{6.10}{-19}{-}{-}
{Incompatible with polyacrylate, Viton$^{\tiny\textregistered}$, natural rubber, silicon rubber and other elastomers.}
{Is considered non-toxic at less than 800 ppm\cite{HoneywellTech}. Hazardous polymerization can occur. If involved in a fire, production under thermal decompose into pyrolysis products containing Hydrogen Fluoride, Carbon Monoxide, Carbonyl halides, and Hydrogen Chloride can occur \cite{Honeywell}.}{1}

\newrow{R1234yf}{114.0}{4.82}{-29}{0/2/2}{ - }
{Incompatible with alkali metals, Zn, Mg and other light metals.}
{If involved in a fire, production under thermal decompose into pyrolysis products containing Fluorine, Carbon Monoxide, Carbonyl halides, and Hydrogen halides can occur. No toxic decomposition should happen under normal conditions.}{1}

\newrow{R1234ze}{114.0}{4.82}{-29}{1/0/0}{ - }{Incompatible with strongly oxidizing materials and finely divided Mg and Al.}
{If involved in a fire, production under thermal decompose into pyrolysis products containing Fluorine, Carbon Monoxide, Carbonyl halides, and Hydrogen halides can occur. Polymerization may also occur.}{1}

\newrow{R13I1}{195.9}{ - }{-22.5}{ - }{ - }
{Incompatible with active metals, fires of hydrides, and materials containing oxygen.}
{Can decompose to Iodine, Hydrogen Fluoride, and Hydrogen Iodide.}{1}

\newrow{R134a}{102.0}{4.320\cite{Praxair}}{-26.5}{1/1/0}{A1}
{Chemically reactive with K, Ca, powdered Al, Mg, Zn. Under high temperature/ high pressure, may react with Al surfaces.}
{Under special circumstances (e.g. high temperature) Carbon monoxide, Carbonyl fluoride, Hydrogen fluoride can be produced. Under normal storage and use, no hazardous decomposition should be produced}{1}

\newrow{R14}{88.0}{3.65}{-128}{0/0/0}{A1}
{Not compatible with aluminium, alloys containing more than 2\% magnesium, alkali metals in powdered form and carbon dioxide above 1000 $^{\circ}$C.}
{If involved in a fire, production under thermal decompose into pyrolysis products containing hydrogen fluorid and carbonyl fluoride.}{1}

\end{tabularx}
\caption{Chemical, physical and compatibility information of the refrigerants (Part 1).}
\label{Table:1}
\end{table}

\begin{table}
\centering
\begin{tabularx}{1\textwidth}{|X|X|X|X|X|X|}

\newrow{\textbf{Refrigerant}}{\textbf{Molecular weight}}{\textbf{Density g/L}}{\textbf{Boiling point degC}}{\textbf{HMIS}}{\textbf{Ashare Safety Group}}{\textbf{Material incompatibility}}{\textbf{Hazardous decomposition products and polymerization}}{0.9}

\newrow{R143a}{84.0}{ - }{-47.6}{-}{A2}
{Can form explosive mixture with air. May react violently with oxidants. Air, Oxidiser.
Non recommended:  Hydrocarbon based lubricant, significant loss of mass by extraction or chemical reaction and Fluorocarbon based lubricant, significant loss of mass by extraction or chemical reaction.}
{Thermal decomposition yields toxic products which can be corrosive in the
presence of moisture.}{1}

\newrow{R125}{120}{ 1.24g/cm3 }{-48.5}{1/1/0}{A1}
{Under very high temperature and/or appropriate pressures freshly abraded aluminum surfaces may cause strongly exothermic reaction. Chemical reactive metals: potassium calcium powdered aluminum, magnesium and zinc. }
{The product is stable. Do not mix with oxygen or air above atmospheric pressure. Any source of high temperatures, such as lighted cigarettes, flames, hot spots or welding may yield toxic and/or corrosive decomposition products. }{1}

\newrow{R22}{86.45}{3 }{-40.1}{1/0/0}{-}
{Chemically reactive metals: potassium, calcium, powdered aluminum, magnesium,
and zinc, powdered metals, powdered metal salts}
{Hazardous decomposition products: Halogens, halogen acids and possibly carbonyl halides. Carbon monoxide, Phosgene, Hydrogen chloride, Hydrogen fluoride, Carbonyl fluoride. }{1}

\newrow{R744}{44}{1.52}{-78.5}{1/0/0}{A1}
{The product is stable under regular conditions}
{ Materials to avoid: strong oxidising agents, strong acids. Hazardous decomposition products: In combustion emits toxic fumes.}{1}

\newrow{R142b}{100.5}{4.18}{-10}{2/4/0}{-}
{Materials to avoid: Light and/or alkaline metals, Alkaline earth metals, Powdered metals, Oxidizing agents, Chlorine, Powdered aluminum, magnesium, zinc, beryllium and their alloys.}
{Hazardous decomposition products: Gaseous hydrogen fluoride (HF)., Gaseous hydrogen chloride (HCl)., Fluorophosgene, Phosgene }{1}

\newrow{R152a}{66.1}{2.738}{-25}{1/4/2}{A2}{Extremely reactive with oxiding materials , such as alkaline, alkaline earth metals, and other reactive chemicals, (i.e. Na, K, Ca, Mg, powdered Al, Zn), brass, and steel . Incompatible with amines, bases, and halogens.}{Under normal condition, hazardous decomposition and/or polymerization products should not be produced. If exposed to fire, hazardous products may be produced.}{1}

\newrow{R218}{188.0}{ 8.17 g/l gas }{ -36.7 }{ - }{A1}
{Stable under normal conditions materials with which gas mixture is incompatible: oxidizing materials and alkali and alkali earth metals. May react violently with chemical active metals as sodium, potassium and barium powdered magnesium, powdered aluminum and organometallics }{Thermal decomposition yields toxical products which can be corrosive in presence of moisture hazardous decomposition products: acid halides}{1}

\newrow{R23}{70.0}{ -2.946 kg /m$^3$}{ -84.4}{ - }{A1}
{ Incompatible materials: metals,polystyrene, natural rubber, alloys of more than 2\% magnesium in the presence of water, nitrosyl fluoride,N$_2$O$_3$, lime at dull red heat, and metals at elevated temperature}
{ Decomposition products: halogenated compounds, oxides of carbon, hydrogen fluoride, thermal decomposition may produce toxic fumes of fluorides. Decomposition products may include the following materials: carbon dioxide carbon monoxide halogenated compounds
carbonyl halides.}{1}

\end{tabularx}
\caption{Chemical, physical and compatibility information of the refrigerants (Part 2).}
\label{Table:2}
\end{table}

\begin{table}
\centering
\begin{tabularx}{1\textwidth}{|X|X|X|X|X|X|}

\newrow{\textbf{Refrigerant}}{\textbf{Molecular weight}}{\textbf{Density g/L}}{\textbf{Boiling point degC}}{\textbf{HMIS}}{\textbf{Ashare Safety Group}}{\textbf{Material incompatibility}}{\textbf{Hazardous decomposition products and polymerization}}{0.9}

\newrow{R290}{44.1}{1.86}{-42}{1/4/0}{A3}
{Incompatible with acids, oxygen, oxidizing materials, copper, some plastics, Chlorine Dioxide.}
{Under normal condition, hazardous decomposition and/or polymerization products should not be produced . May produce carbon monoxide and other toxic gasses under thermal decomposition.}{1}

\newrow{R32}{52.0}{11.4}{-51.7}{1/4/1}{A2}
{incompatible with acids and oxidizing materials as Na, K, Ca, Zn, Mg, powdered Al, and other active metals. Incompatible with air and moisture.}
{No hazardous decomposition/polymerization should be produced under normal conditions.}{1}

\newrow{R600a}{58.1}{8.93}{-11.7}{1/4/0}{A3}
{Incompatible with oxiding materials, halogenated hydrocarbons, halogens, and metal catalysts.}
{No hazardous decomposition/polymerization should be produced under normal conditions. May produce carbon monoxide and other toxic gasses under thermal decomposition.}{1}

\newrow{R7146}{146.1}{6.17}{-63.7}{1/0/0}{ - }
{Stable with most chemical, except metals other than aluminium, stainless steel, copper brass, silver, at elevated temperatures (>204$^{\circ}$C). Also reacts violently with disilane.}
{Decomposes into Sulfur oxides and hydrogen fluorine.}{1}

\end{tabularx}
\caption{Chemical, physical and compatibility information of the refrigerants (Part 3).}
\label{Table:3}
\end{table}
%

%
\section{Estimation of Gas Parameters}
\subsection{Stopping power} \label{sec:Spower}
\begin{figure}
\centering
\includegraphics[width=1.1\columnwidth]{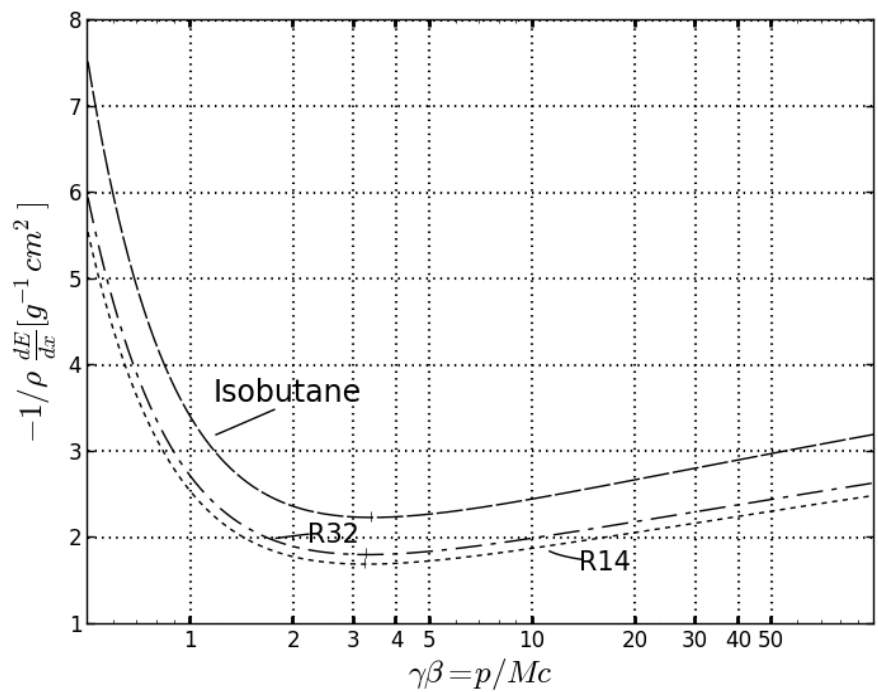}
\caption{Energy loss as a function of the relativistic time dilation factor $\gamma\beta$ for various refrigerants.}
\label{fig:spower}
\end{figure}

Quantities such as the minimum ionization energy can be found if the stopping power is known. An approximate expression for moderately relativistic particles in the momentum region $0.1\leq \beta\gamma=p/Mc \leq 1000$ can be found using the Bethe-Bloch equation, given by \cite{PDG}
\begin{equation}
\dfrac{1}{\rho}\left<-\dfrac{dE}{dx}\right>=Kz^2\dfrac{Z}{A}\dfrac{1}{\beta^2}\left[\dfrac{1}{2}\ln\dfrac{2m_ec^2\beta^2\gamma^2T_{max}}{I^2}-\beta^2-\dfrac{\delta(\beta\gamma)}{2}\right]
\end{equation}
where $\left<-\dfrac{dE}{dx}\right>$ is the mean energy loss per length, $\rho$ is the density of the medium,  $I$ is the mean excitation energy, and $\delta(\beta\gamma)$ is the density effect correction function to ionization energy loss. $K$ is a constant given by $4\pi N_A r_e^2m_ec^2$, and $T_{max}$ is the maximum energy transfer in a single collision, given by
\begin{equation}
T_{max}=\dfrac{2m_ec^2\beta^2\gamma^2}{1+2\gamma m_e/M +(m_e/M)^2},
\end{equation}
where M is the mass of the incoming particle.\\
The mean excitation energy $I$ for a composite medium can be approximated from the composite atoms by the relation\cite{seltzer1982a}.
\begin{equation}
I=\exp\left\lbrace\left[\sum_jw_j(Z_j/A_j)\ln I_j\right]/\left<Z/A\right>\right\rbrace
\end{equation}
where $w_j$, $Z_j$, $A_j$ and $I_j$ is the fraction by weight, atomic number, atomic weight and mean ionization energy, respectively, of the $j$'th constituent. The shape of the $\delta (\gamma\delta)$ function for non-conducting materials can be approximated by
\begin{equation}
	\delta(\gamma\delta)=\left\{ \begin{array}{ll}
		2(\ln 10)X-\bar{C} & \text{if } X\geq X_1;\\
		2(\ln 10)X-\bar{C} & \text{if } X_0\leq X < X_1;\\
		0 & \text{if } X < X_0;
		\end{array}
	\right.
\end{equation}
where $X=\log_{10}(\gamma\delta)$. For finding an approximate expression for the parameters $\bar{C}$, $X_0$ and $X_1$ based on experimental fits, we refer to 
\cite{Sternheimer:1971zz}.
 For gases with  momenta below $\beta\gamma$, the density effect correlation function can  be neglected. A plot of the calculated energy loss for different refrigerants is show in figure \ref{fig:spower}.

\subsection{Radiation length}
The radiation length of an atom can be found by \cite{Tsai:1973py}
\begin{equation}
X_{0}=716.405(\text{cm}^{-2}\text{mol}) A/\left[Z^2(L_{rad}-f(z))+ZL_{rad}'\right]
\end{equation}
\begin{equation}
f(z)=z^2\sum_{n=1}^{\infty}\dfrac{1}{n(n^2+z^2)}\approx 1.202z-1.0369z^2+\dfrac{1.008z^3}{1+z}
\end{equation}
where $L_1$ and $L_2$ is given by table \ref{table:L}, $f(z)$ is the one-photon exchange approximation, and $z=\alpha Z$, $\alpha$ being the fine-structure constant and $Z$ is the atomic number. This formula, however, only holds for free atoms. When one wants to find the stopping power for a molecule, one has to take into account the influence from molecular bindings, crystal structures and polarization of the medium. By neglecting these effects, however, one can find an approximate expression by weighting the radiation length of the single atoms
\begin{equation}
\dfrac{1}{X_0} = \dfrac{1}{A_{molecule}}\sum_j \dfrac{A_j}{X_{0j}},
\label{eq:weightedlength}
\end{equation}
where $j$ reffers to the $j$'th atom's constituent.

\begin{table}
\centering
\begin{tabular}{c c c c}
Z & $L_1$ & $L_2$ \\
\hline
1 & 5.31 & 6.144 \\
2 & 4.79 & 5.621 \\
3 & 4.74 & 5.805 \\
4 & 4.71 & 5.924 \\
5$\geq$ & $\ln ( 184.15Z^{-1/3})$ & $\ln (1994Z^{-2/3})$
\end{tabular}
\label{table:L}
\caption{$L_1$ and $L_2$ expressions for various atom numbers.}
\end{table}

\subsection{Estimation of ionization pair production}
In order to model the number of primary ionization caused by a single particle, the cross section for all the particle-atom interaction should be calculated. The number of primary electrons per unit length would then be the integral over energy across all the energy transfer cross sections. This is tedious work since all electron orbital transfers have to be considered. An easier, but approximate correlation between primary ionization and atom number has been found based on experimental data by \cite{Smirnov:2005yi}
\begin{equation}
N_P=3.996\dfrac{Z_m}{\bar{Z}^{0.4}}-0.025\left(\dfrac{Z_m}{\bar{Z}^{0.4}}\right) \text{cm}^{-1},
\end{equation}
which holds for normal pressure and temperature (NPT) (1 atm, 20$^{\circ}$C). For different pressure and temperature, the number scales with the density. This value should only be taken as a rough estimation though. This formula has proven to work best for hydrocarbons and worst for molecules consisting mainly of fluorine, differing as much as 30\% from the experimental value for CF$_4$.
\par
The total number of ionization has proven to be more difficult to estimate. Whereas no general formula has been derived, the most straightforward method will be to use the cross sections used to calculate the primary ionization electrons, and use Monte Carlo simulations to track the production of secondary electrons from primary electrons. The total number of pair ionization turns out to be dependent on the incoming particle energy and mass, and a general expression can therefore be difficult to find. For an incoming particle, $W=\dfrac{\Delta E}{N}$ defines the average energy cost to produce an ion pair. Luckily, this turns out to be a function slowly varying with particle energy\cite{ICRU}, and can therefore be taken to be a constant in an energy interval. The total ionization per unit length can then be found by
\begin{equation}
N_{T}=\rho\dfrac{dE}{dx}\dfrac{1}{W}
\end{equation}
If the $W$ values for specific gases are know, the average $W$ value for a gas mixture can be found by\cite{Smirnov:2005yi}
\begin{equation}
\bar{W}=\sum_m\left[f(n_m)Z(n_m)W(n_m)\right]\left/ \sum_m\left[f(n_m)Z(n_m)\right]\right. ,
\end{equation}
where $n_m$ denotes the index of the molecule, and $f(n_m)$ denotes the relative number of molecules of the given sort in the mixture.\\
The value of W is difficult to predict, and there is not a direct way to give a proper estimate based on experimental data alone. A montecarlo simulation is in preparation which uses  the photoabsorption ionization and relaxation (PAIR) model \cite{Smirnov:2005yi} and it will be the subject of a forthcoming paper.
%
%
%
%
\section{Quantum chemistry calculations} \label{sec:Quantum}
In this section of the paper we describe the calculation of gas properties using the quantum chemistry calculation tool \nwchem. The ionization energies, electron affinities and the absorption spectrums are calculated for several kinds of potential Eco-friendly gas replacements. The results are validated and compared with different methods or calculation schemes. In the last part, the problems related to the  computation of the Townsend parameter and prop sects for future works are discussed. 
\par
Developments of quantum chemistry in the past few decades have improved the calculation to be accurate enough to be compared with real experiments. These calculations enable us to study gas properties from atom level, where the underlying theory is believed to be more fundamental and may therefore provide us more general descriptions of the chemical properties. 
\par
In the paper, we used the \nwchem  software \cite{Valiev20101477}  for computation, and the \pymol \cite{PyMOL} tool for visualization. The calculation determines the electronic structure of the molecule system and can provide basic information of the system like ground state energy, dipole momentum, charged density and molecular orbit. To obtain properties more suitable to be compared to experimental data, the software also provides frameworks to compute the absorption spectrum and the ionization energy. 
\par
Here we first show  the basic geometrical information for interested molecules. Then we describe the procedure to calculate the absorption spectrum and in the following we present the methods to calculate the ionization energy and the electron affinity. In the end, we discussed the possibility to estimate the Townsend parameter under some certain environment of particle detectors. 
\subsection{Molecules and their optimized geometries} \label{sec:Molecules}
\begin{figure}[]
  \begin{center}
    \subfigure[$CH_4$]{
    \includegraphics[width=0.23\textwidth]{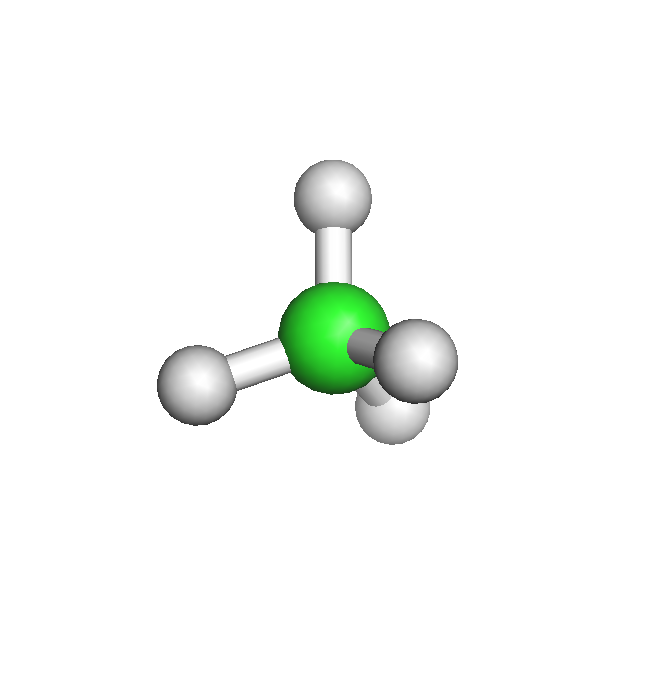}
  }
    \subfigure[$CF_4$]{
    \includegraphics[width=0.23\textwidth]{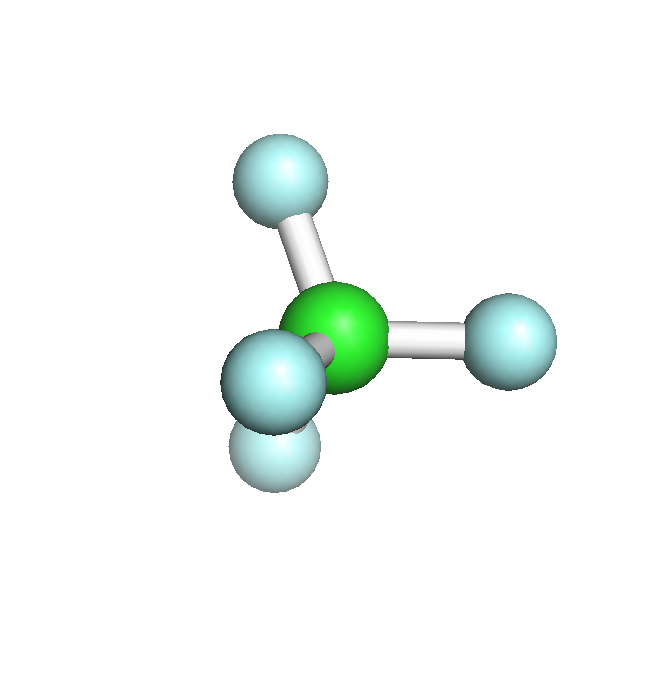}
  }
    \subfigure[R12$(CCl_2F_2)$]{
    \includegraphics[width=0.23\textwidth]{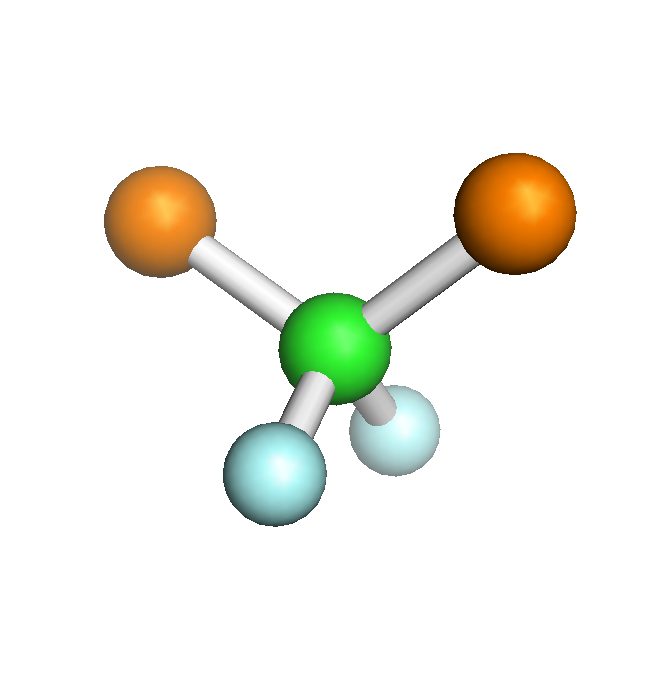}
  }
    \subfigure[CF3I$(CF_3I)$]{
    \includegraphics[width=0.23\textwidth]{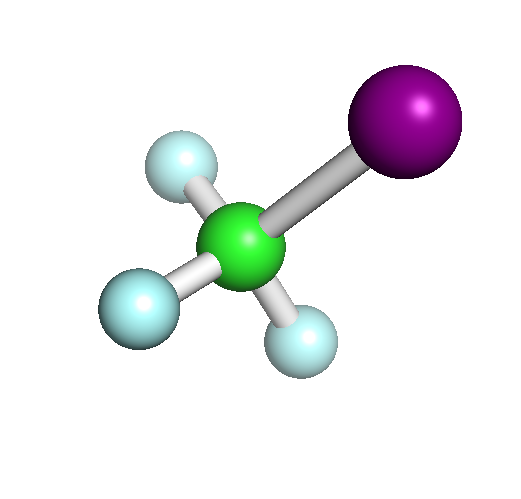}
  } \\
    \subfigure[R134a($CH_2FCF_3$)]{
    \includegraphics[width=0.23\textwidth]{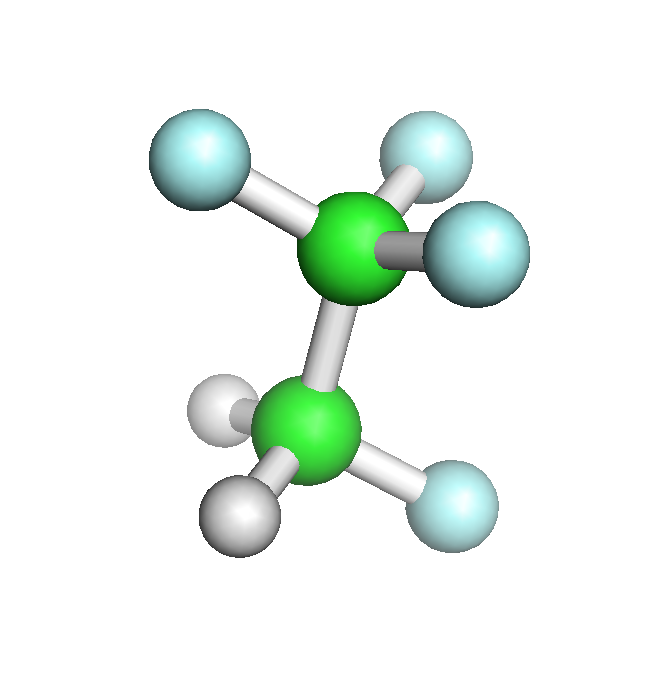}
  }  
    \subfigure[R152a($C_2H_4F_2$)]{
    \includegraphics[width=0.23\textwidth]{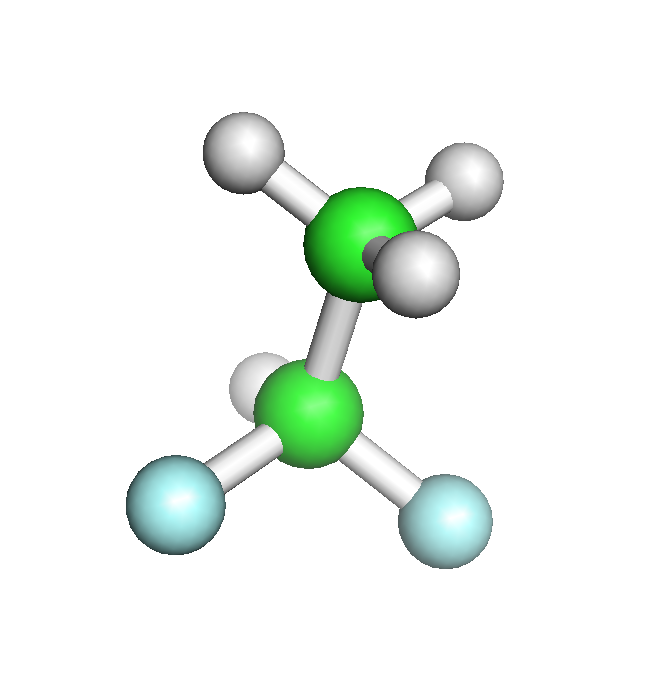}
  } \\
    \subfigure[HFO1234ze($CFHCHCF_3$)]{
    \includegraphics[width=0.23\textwidth]{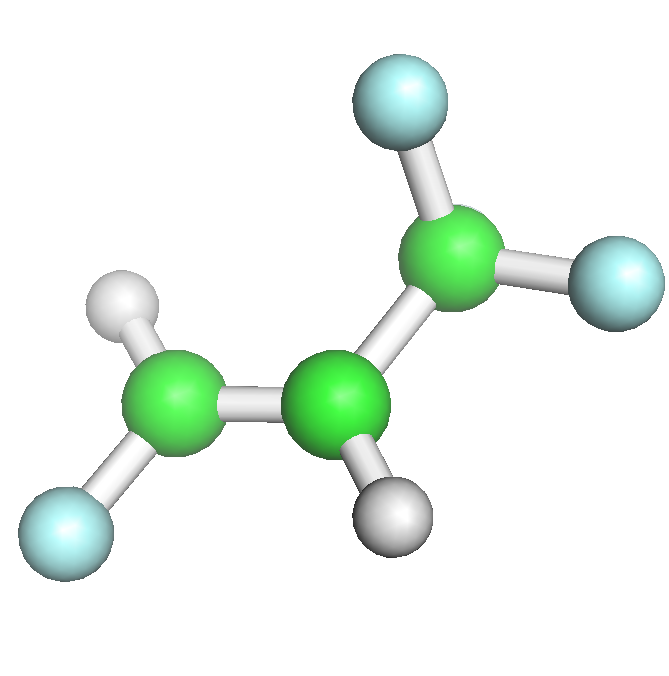}
  } \\
    \subfigure[HFO1234yf($CH_2CFCF_3$)]{
    \includegraphics[width=0.23\textwidth]{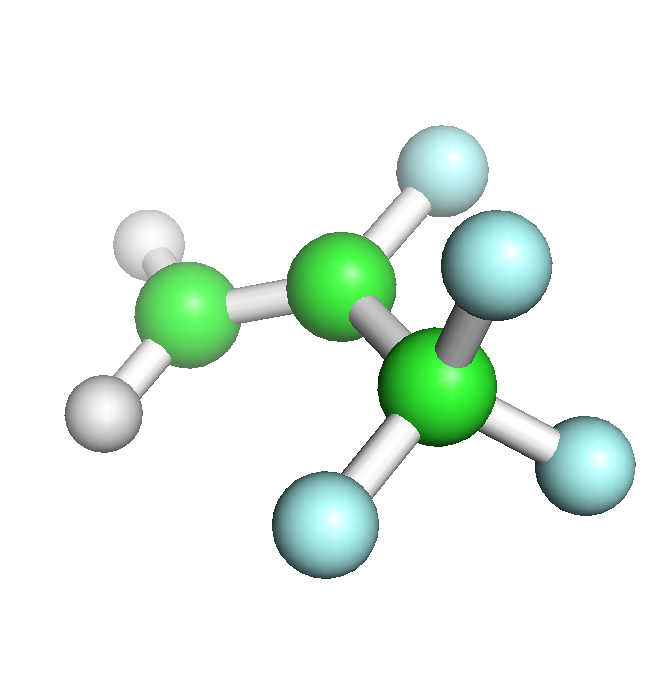}
  } \\
    \subfigure[HFO1233zd($CHClCFCF_3$)]{
    \includegraphics[width=0.23\textwidth]{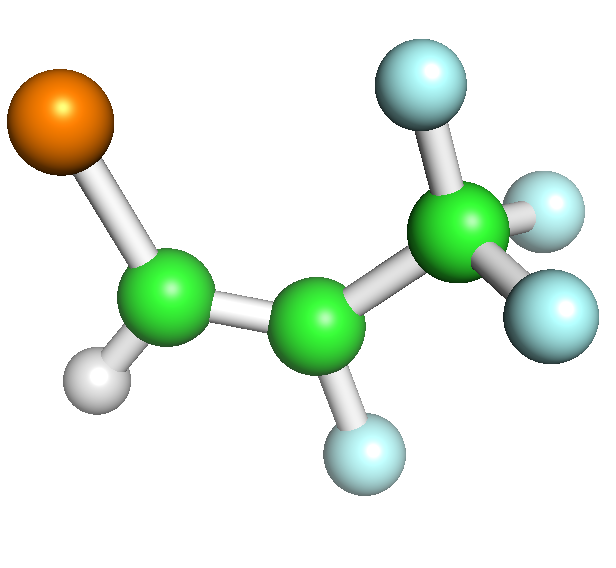}
  }
    \caption{ Optimized ground state geometries of gas molecules. }
    \label{fig:geometry}
  \end{center}
\end{figure}

\begin{figure}[]
  \begin{center}
    \subfigure[$CH_4$]{
    \includegraphics[width=0.23\textwidth]{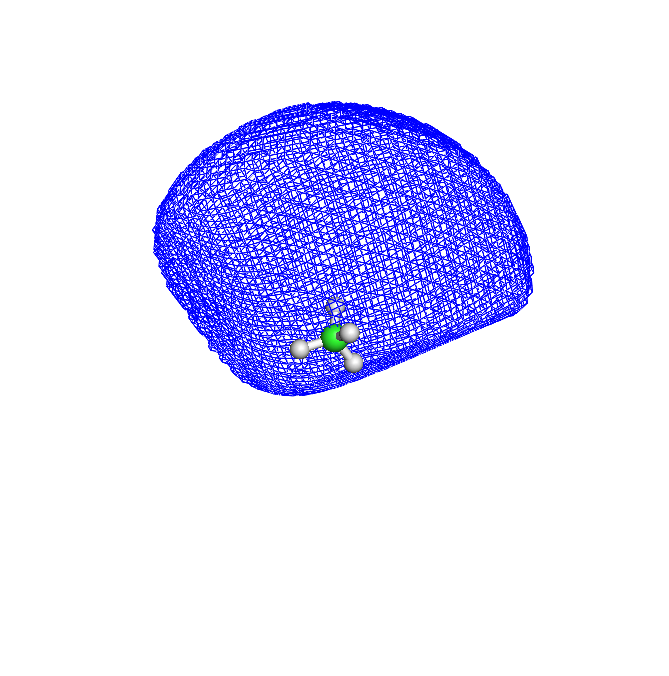}
  }
    \subfigure[$CF_4$]{
    \includegraphics[width=0.23\textwidth]{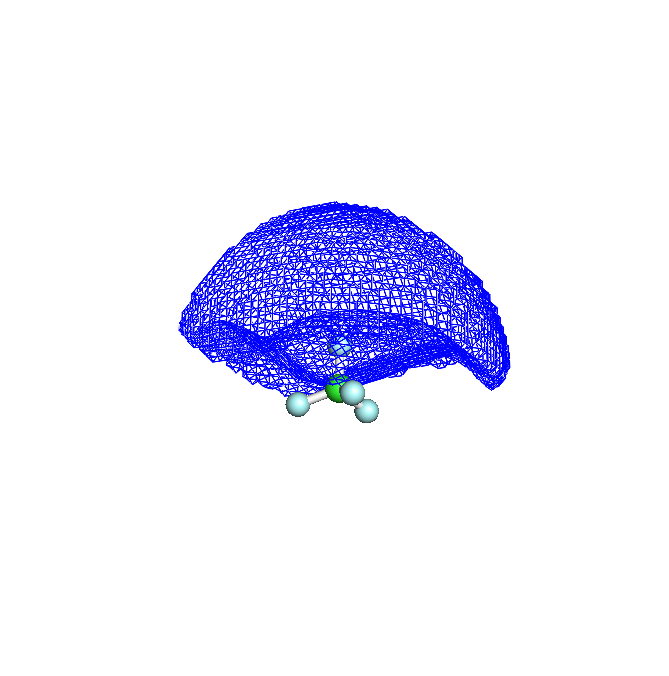}
  }
    \subfigure[R12$(CCl_2F_2)$]{
    \includegraphics[width=0.23\textwidth]{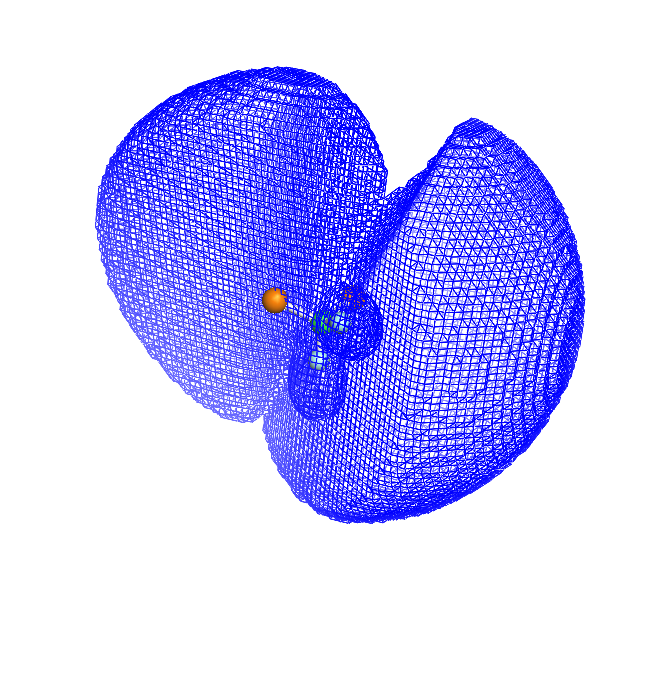}
  }
    \subfigure[CF3I($CF_3I$)]{
    \includegraphics[width=0.23\textwidth]{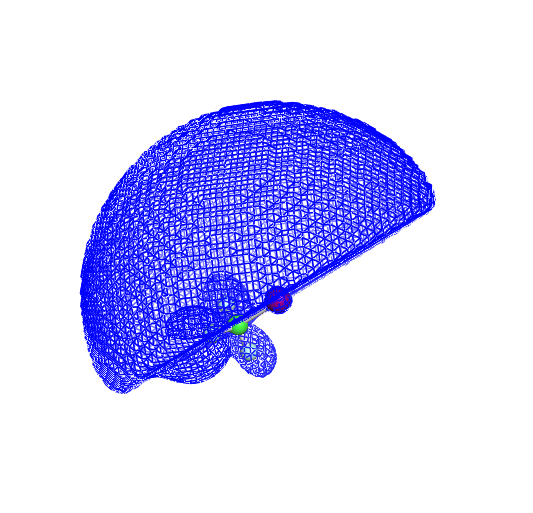}
  } \\ 
    \subfigure[R134a($CH_2FCF_3$)]{
    \includegraphics[width=0.23\textwidth]{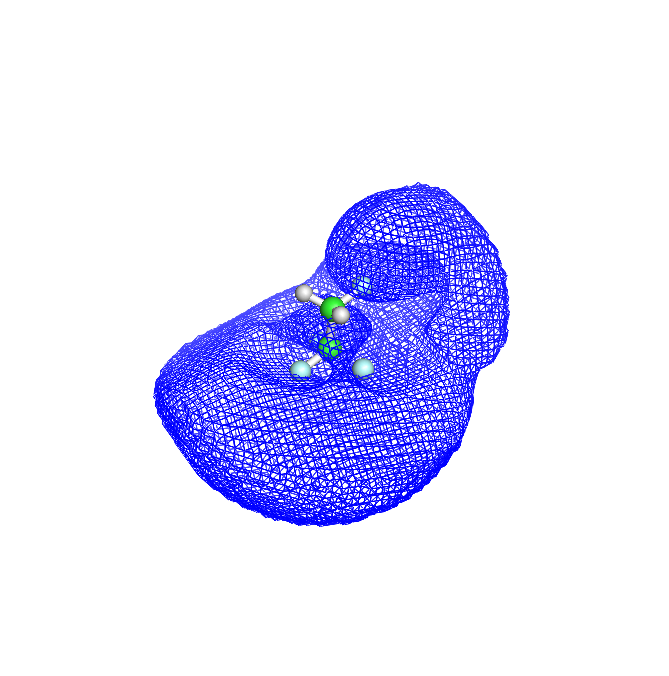}
  } 
    \subfigure[R152a($C_2H_4F_2$)]{
    \includegraphics[width=0.23\textwidth]{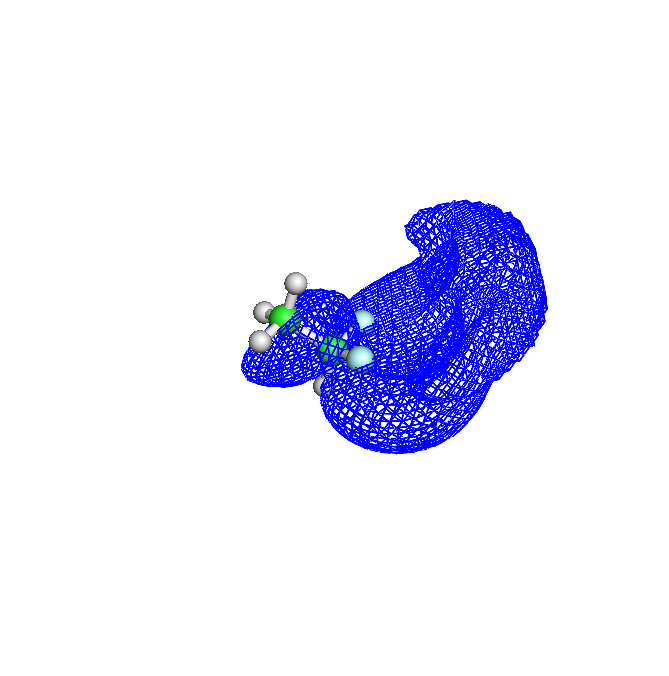}
  } \\
    \subfigure[HFO1234ze($CFHCHCF_3$)]{
    \includegraphics[width=0.23\textwidth]{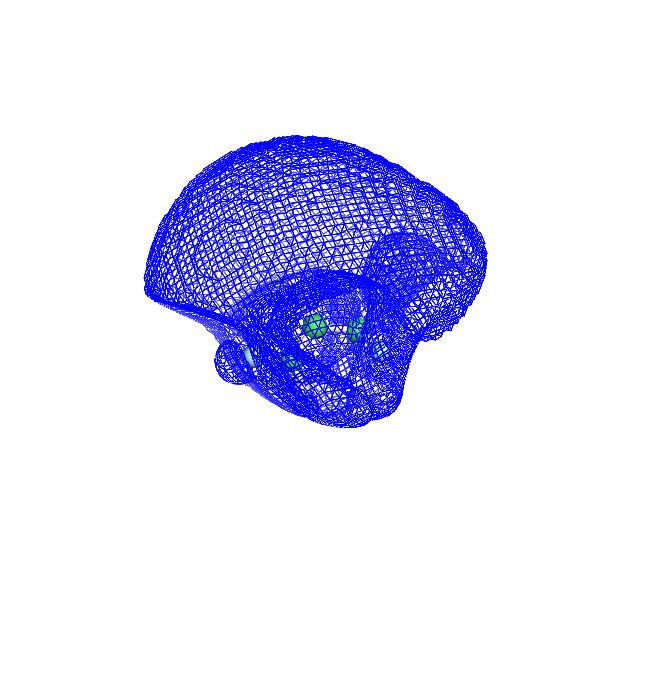}
  } \\
    \subfigure[HFO1234yf($CH_2CFCF_3$)]{
    \includegraphics[width=0.23\textwidth]{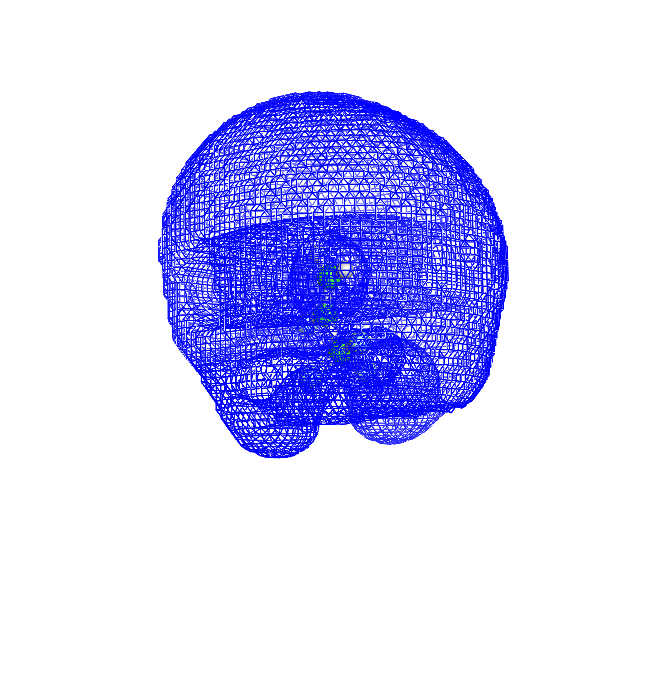}
  } \\
    \subfigure[HFO1233zd($CHClCFCF_3$)]{
    \includegraphics[width=0.23\textwidth]{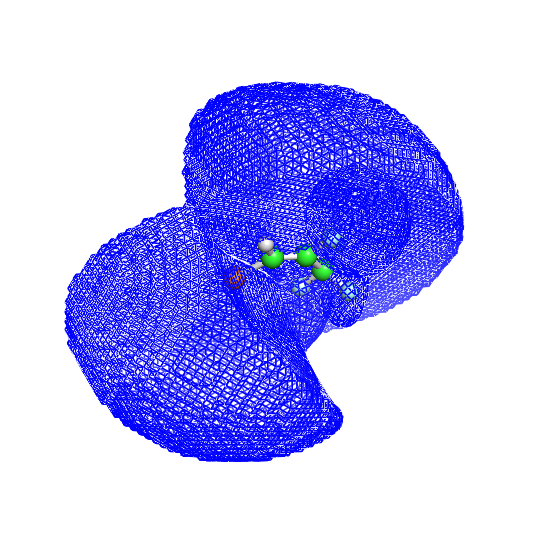}
  }
    \caption{ Highest Occupied Molecular Orbitals (HOMO) of gas molecules. }
    \label{fig:HOMO}
  \end{center}
\end{figure}

The freon gases we have been interested in are R134a ($CH_2FCF_3$), R152a ($C_2H_4F_2$),
HFO1234ze ($CFHCHCF_3$), HFO1234yf  ($CH_2CFCF_3$), $CF_3I$ and 
HFO1233zd ($CHClCFCF_3$). Meanwhile we choose R12$(CCl_2F_2)$ to be the standard freon 
gas model. $CH_4$ and $CF_4$ are the molecules we were used to make a comparison 
between the \nwchem   calculated results and experimental results, in order to check the 
reliability of \nwchem. Fig.(\ref{fig:geometry}) shows the optimized ground state 
geometries of gas molecules, where the green balls stand for carbon atoms, the grey 
balls stand for hydrogen atoms, the indigo balls stand for fluorine atoms, the brown 
balls stand for chlorine atoms, and purple ball for iodine. Fig.(\ref{fig:HOMO}) shows 
the highest occupied molecular orbitals (HOMO) of gas molecules.

\section{Calculation of absorption spectrum}

The excitation energy of a molecule is one of the fundamental properties of molecular interactions one can get from experiment. To study these properties, we used the framework introduced in Ref.\cite{doi:10.1021/ct200137z} to simulate the time-dependent response of molecules under external fields using quantum chemical calculations. 
The framework implemented is the Real-Time Time-Dependent Density Functional Theory (RT-TDDFT)~\cite{:/content/aip/journal/jcp/127/15/10.1063/1.2790014} method within the \nwchem, making it capable of doing the simulation beyond small perturbation from the ground state. 

In our case, we are mostly interested in low excitations of small molecules. We therefore adopt the procedure as described in section~3 of the reference~\cite{doi:10.1021/ct200137z}. The choice of external field is $\delta$-function-like electric field ``kick''

\begin{equation}
\mathbf{E}(t) = \kappa \cdot exp\left[-(t-t_0)^2/2w^2\right]\hat{d},
\end{equation}
where $t_0$ is the center of pulse, $w$ is the pulse width, which has dimensions of time, $\hat{d} = \hat{x}, \hat{y}, \hat{z}$ is the polarization of the pulse, and $\kappa$ is the maximum field strength.

The system is then evolved in time, and the dipole moment can be calculated with respect to the added dipole coupling term

\begin{equation}
\mathbf{V}^{app}_{\mu\nu}(t) = - \mathbf{D}_{\mu\nu} \cdot \mathbf{E}(t),
\end{equation}
where $\mathbf{D}$ is the transition dipole tensor of the system.
Then we Fourier transform the dipole signals to construct the complex polarizability tensor $\alpha_{i,j}(\omega)$, and finally the dipole absorption spectrum is 
\begin{equation}
S(\omega) = \dfrac{1}{3} Tr\left[\sigma(\omega)\right] = \dfrac{4\pi\omega}{3c}Tr\left[Im\left[\alpha_{i,j}(\omega)\right]\right].
\end{equation}

We first validate our calculation by repeating the calculation of $CH_4$ lowest 
excitation energy in Table~(\ref{tab:IE}) of the paper~\cite{doi:10.1021/ct200137z}. With the same basis set (6-311G) and functional (B3LYP), our calculation gives 11.16 eV excitation energy, which is consistent with the 11.13 eV in the paper. The small difference can come from other minor uncertainty sources like the choice of time separation in the simulation.  

\begin{figure}[]
  \begin{center}
    \subfigure[X-dipole, LDA]{
    \includegraphics[width=0.31\textwidth]{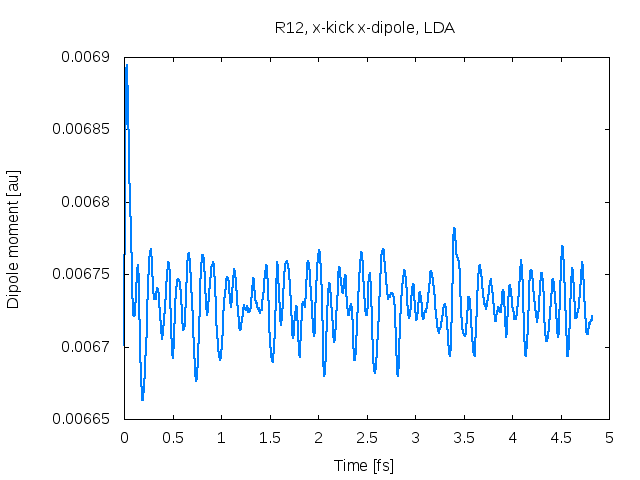}
  }
    \subfigure[Y-dipole, LDA]{
    \includegraphics[width=0.31\textwidth]{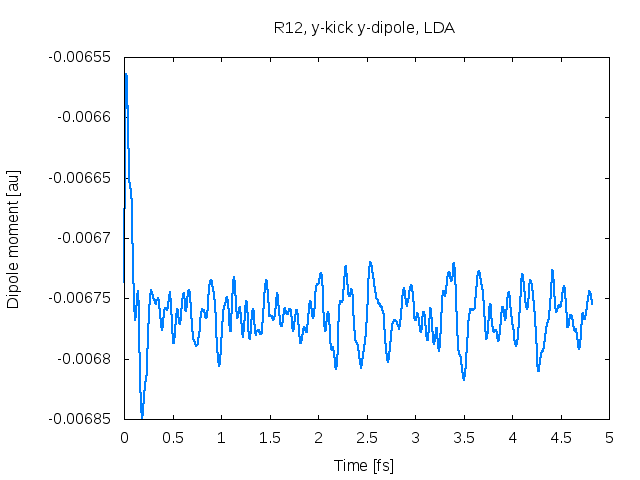}
  } 
    \subfigure[Z-dipole, LDA]{
    \includegraphics[width=0.31\textwidth]{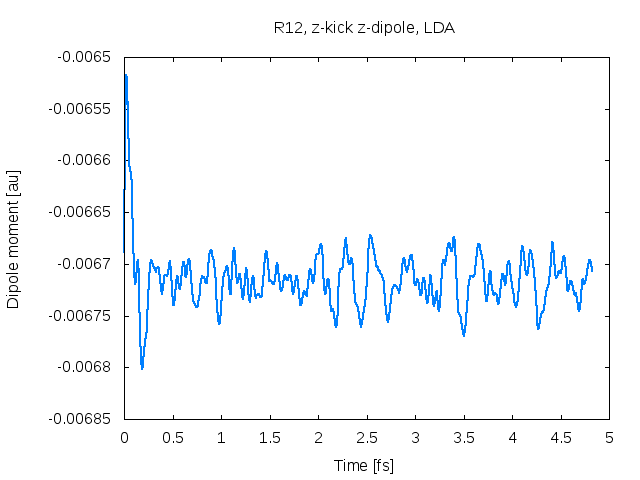}
  }\\
    \subfigure[X-dipole, B3LYP]{
    \includegraphics[width=0.31\textwidth]{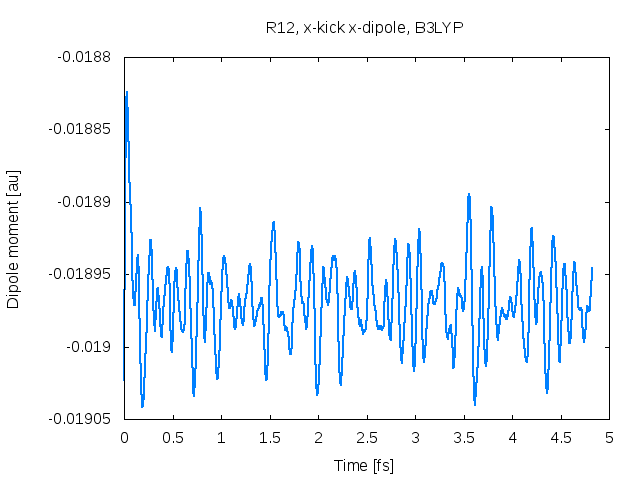}
  }
    \subfigure[Y-dipole, B3LYP]{
    \includegraphics[width=0.31\textwidth]{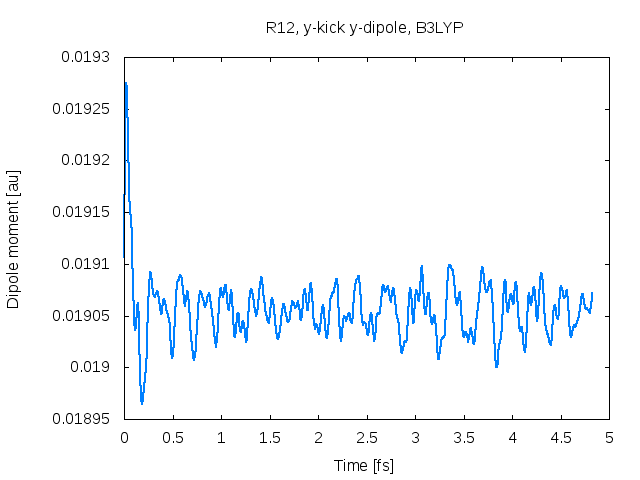}
  }
    \subfigure[Z-dipole, B3LYP]{
    \includegraphics[width=0.31\textwidth]{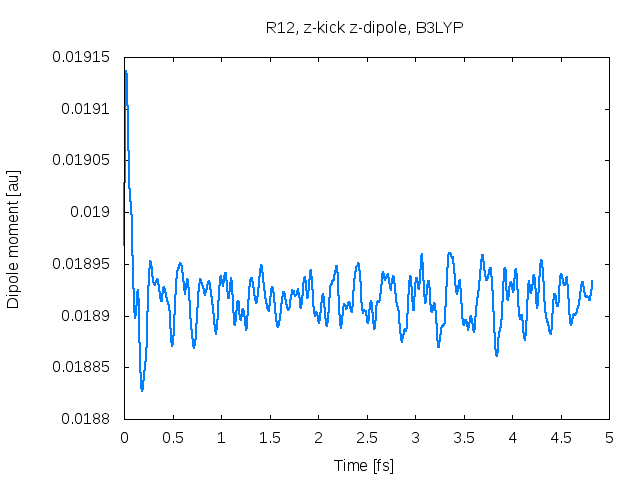}
  }
    \caption{ R12$(CCl_2F_2)$ dipole moment in the 3 kicked directions, using the 6-31~G* basis and LDA (top) and B3LYP (bottom) functionals. }
    \label{fig:r12dipole}
  \end{center}
\end{figure}

\begin{figure}[htbp]
  \centering
  \includegraphics[width=0.5\textwidth]{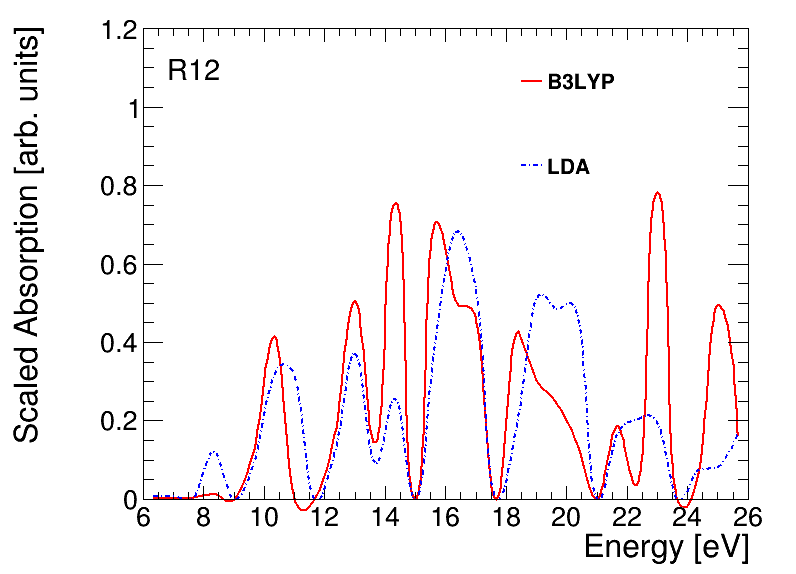}
  \caption{\label{fig:comp_functional} A comparison of R12$(CCl_2F_2)$ absorption spectrum using different functionals. }
\end{figure}

We further checked the dependence of wave function basis sets as well as density functional of our results. The dependence of basis sets is found to be much smaller than the functional one. Fig.(\ref{fig:r12dipole}) and Fig.(\ref{fig:comp_functional}) compares the results using two different functionals. As shown in the figures, although the strength varies large between the two functionals, the excitation energies are similar, especially the lowest one, which is also the same story as in the reference~\cite{doi:10.1021/ct200137z}.

\begin{figure}[]
  \begin{center}
    \subfigure[$CH_4$]{
    \includegraphics[width=0.31\textwidth]{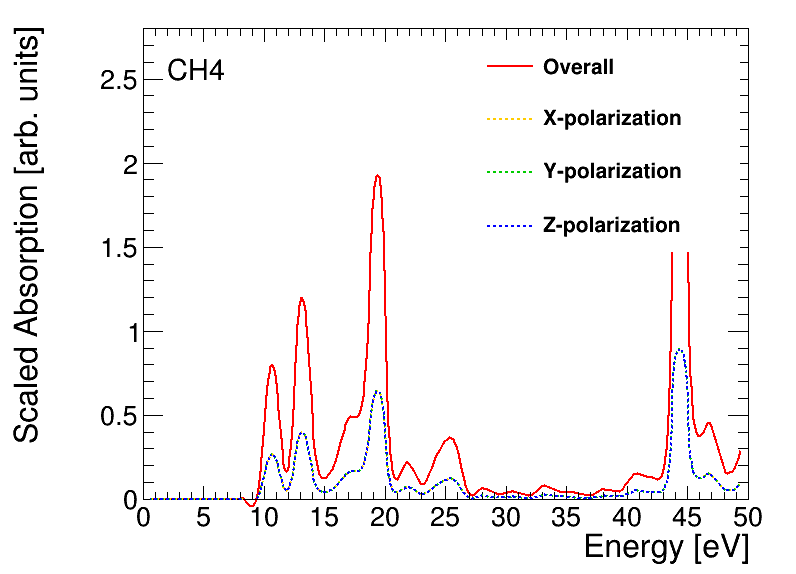}
  }
    \subfigure[$CF_4$]{
    \includegraphics[width=0.31\textwidth]{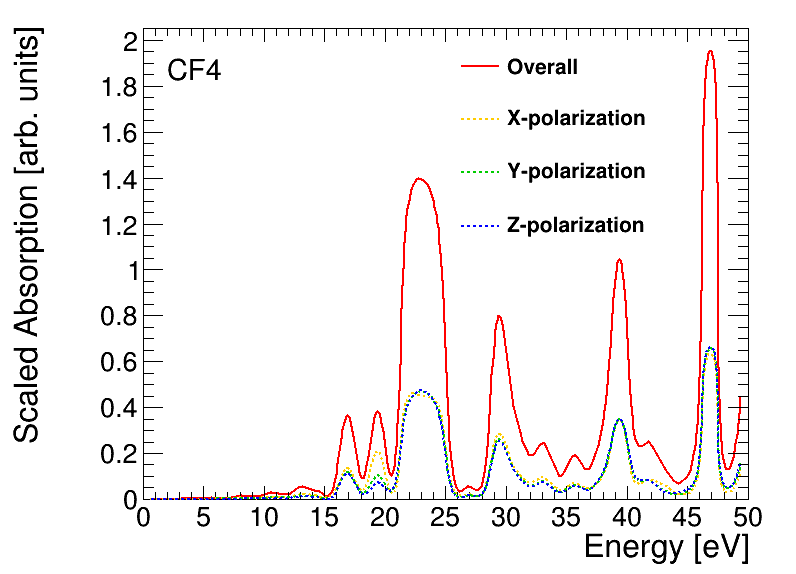}
  } 
    \subfigure[$R12(CCl_2F_2)$]{
    \includegraphics[width=0.31\textwidth]{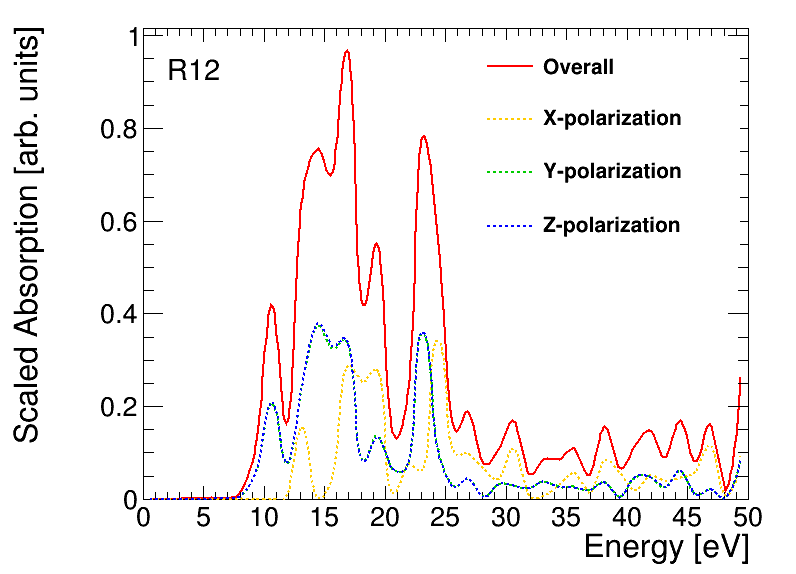}
  } \\
    \subfigure[$CF_3I$]{
    \includegraphics[width=0.31\textwidth]{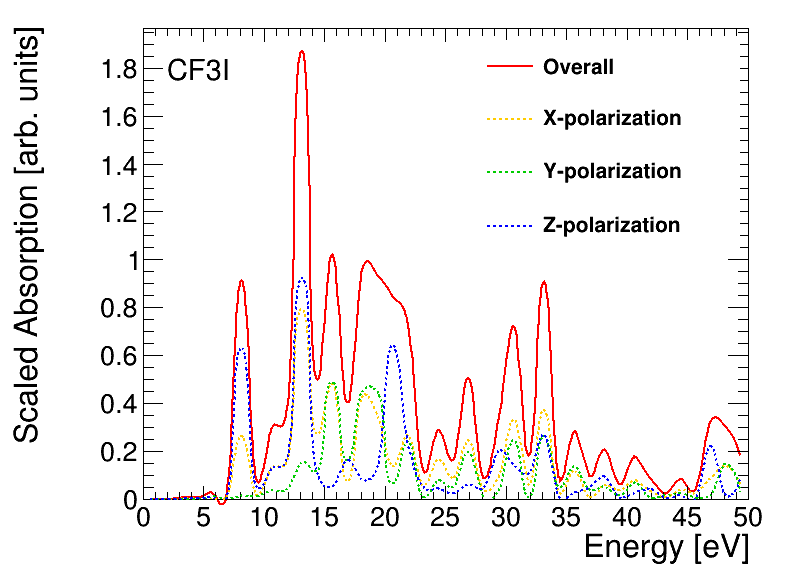}
  } 
    \subfigure[R134a($CH_2FCF_3$)]{
    \includegraphics[width=0.31\textwidth]{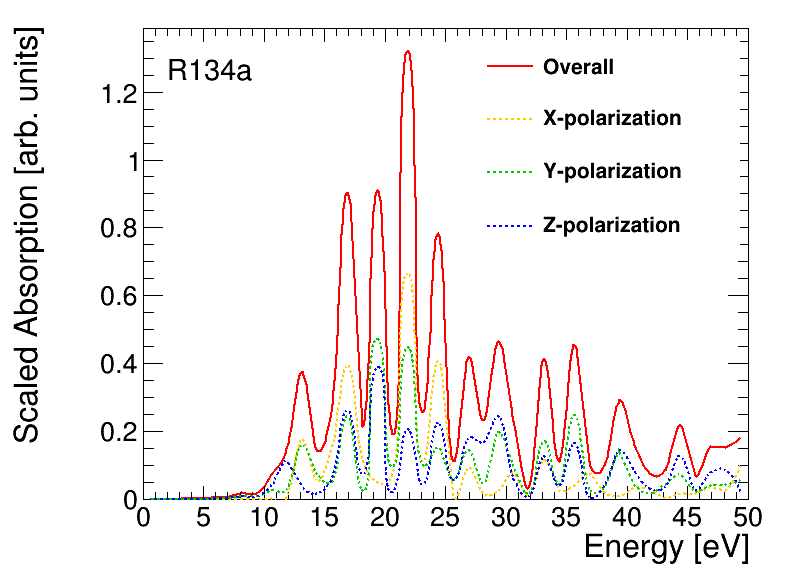}
  } 
    \subfigure[R152a($C_2H_4F_2$)]{
    \includegraphics[width=0.31\textwidth]{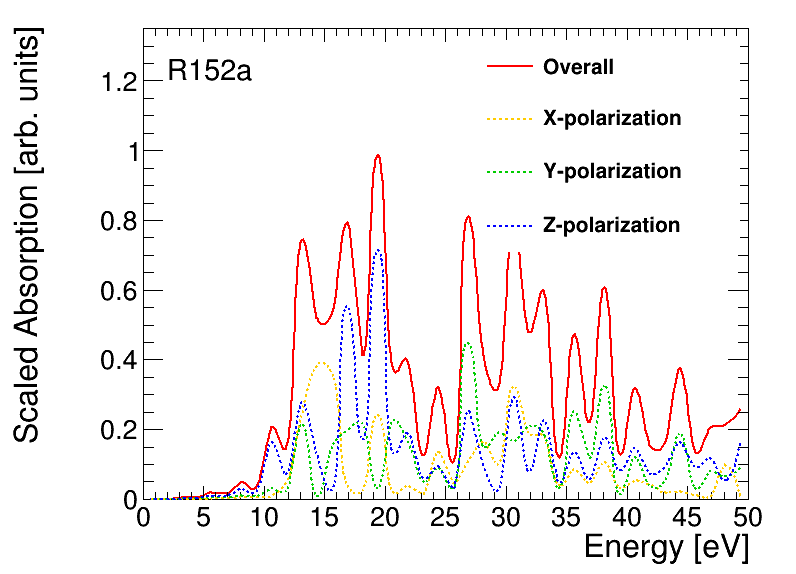}
  } \\
    \subfigure[HFO1234ze($CFHCHCF_3$)]{
    \includegraphics[width=0.31\textwidth]{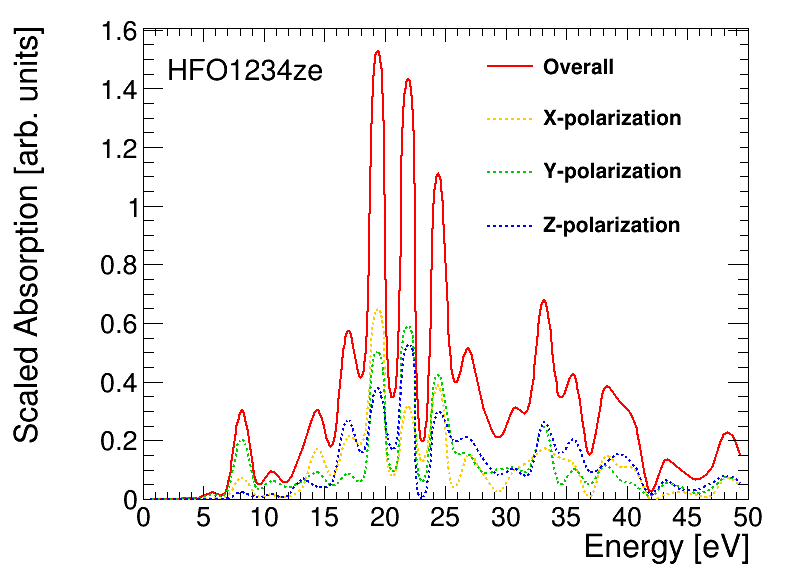}
  }
    \subfigure[HFO1234yf($CH_2CFCF_3$)]{
    \includegraphics[width=0.31\textwidth]{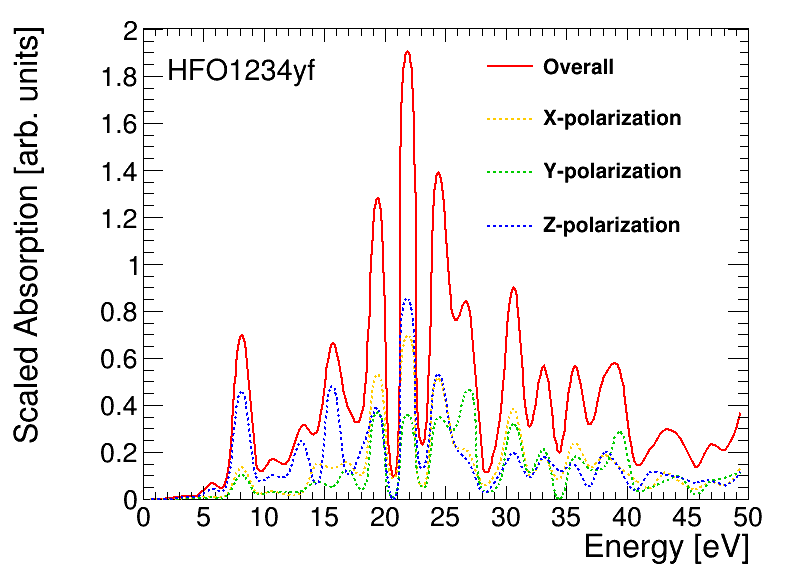}
  } 
    \subfigure[HFO1233zd($CHClCFCF_3$)]{
    \includegraphics[width=0.31\textwidth]{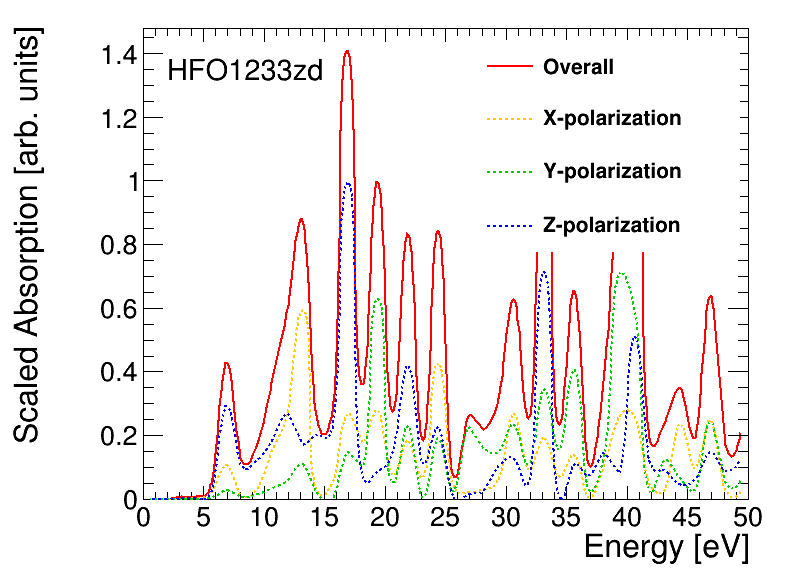}
  } 
    \caption{ Absorption spectrum of interested gas molecules. The calculation is performed in the AO basis, where the solid line corresponds to the overall dipole strength and the dashed lines correspond to polarizations in the coordinate basis directions. }
    \label{fig:absorption}
  \end{center}
\end{figure}

The same calculation is performed to gas molecules of our interest. The optical absorption spectrums are compiled in Fig.(\ref{fig:absorption}). 

\begin{figure}[]
  \begin{center}
    \subfigure[R12 and HFO Gases]{
    \includegraphics[width=0.45\textwidth]{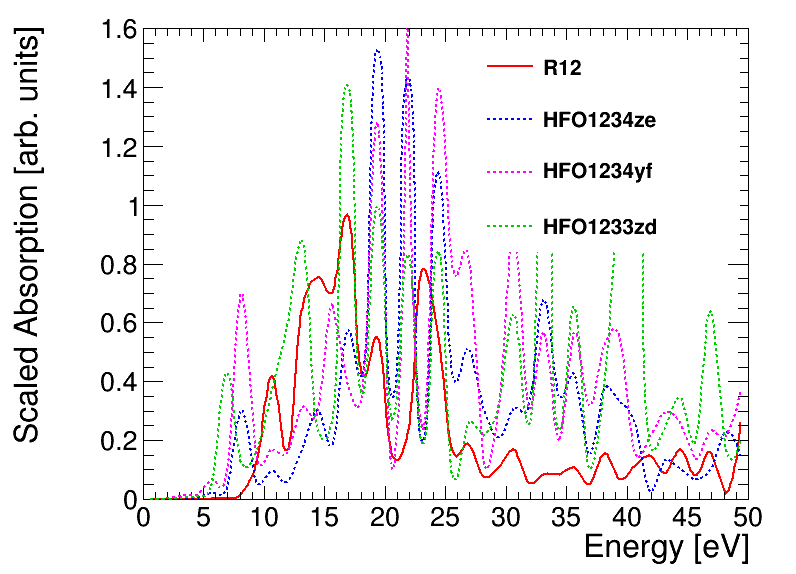}
  }
    \subfigure[R12 and other Freon Gases]{
    \includegraphics[width=0.45\textwidth]{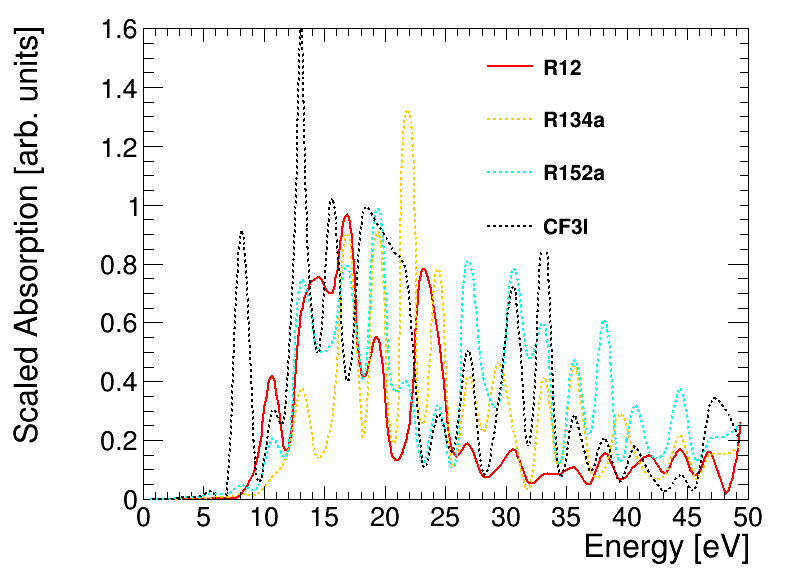}
  } \\
    \subfigure[All Freon Gases]{
    \includegraphics[width=0.45\textwidth]{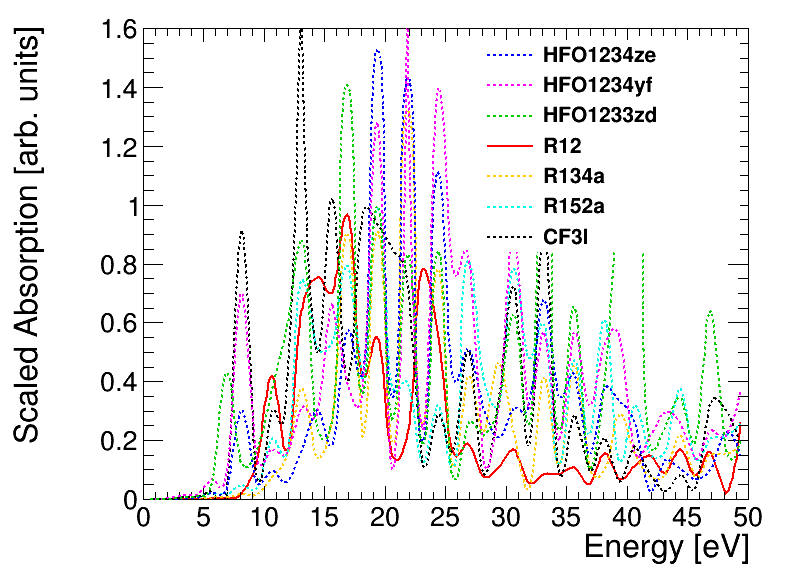}
  } 
    \caption{ A comparison of absorption spectrum among molecules. }
    \label{fig:compabs}
  \end{center}
\end{figure}

In the Fig.(\ref{fig:compabs}) we put together all spectra of interesting molecules. Comparing to the R12, which is one of the Freon gas spectrum, we found the similarity between molecules are in good agreement with their structures. In other words, the absorption spectrum is closer  to that of R12 if the candidate molecule is similar to the R12 molecular structure. 

\section{Calculation of the vertical ionization energy and the electron affinity}

The vertical ionization energy and the electron affinity are another two important parameters closely related to experiments. In the case of small molecules, there is a simple way to do the estimation. The Koopman's theorem states that at in closed-shell Hartree-Fock (HF) theory, the negative value of the Highest Occupied Molecular Orbital (HOMO) energy is a good approximation of to the experimental ionization energy (IE). Similarly, it also suggests that the electron affinity (EA) is equal to the negative of the Lowest Unoccupied Molecular Orbital (LUMO) energy. However, the EA part of the prediction is in general not reliable because of the large effect of orbital relaxation on the LUMO eigenvalue~\cite{:/content/aip/journal/jcp/113/15/10.1063/1.1308547,doi:10.1021/jp061633o}. 

In this section, besides estimating the vertical ionization energy from the HOMO, we present a more direct way to do the calculation. In this calculation, we optimized the geometry of both the neutral gas molecule and the positive one charged ion accordingly, for the spin-Unrestricted Hartree-Fock (UHF) wave functions. Then the vertical ionization energy is calculated from the ground state energy of positively charged ion by subtracting that of  the neutral molecule. The results are shown in Table.(\ref{tab:IE}).

For the electron affinity calculation, we calculated the LUMO with two methods. The Hartree-Fock method, computes the closed-shell Restricted Hartree-Fock (RHF) wave functions, and the Density Functional Theory (DFT), computes the closed shell densities and Kohn-Sham orbitals in the Generalized Gradient Approximation. The results are also shown in Table.(\ref{tab:EA}).

\begin{figure}[]
  \begin{center}
    \subfigure[R12 HOMO and LUMO]{
    \includegraphics[width=0.41\textwidth]{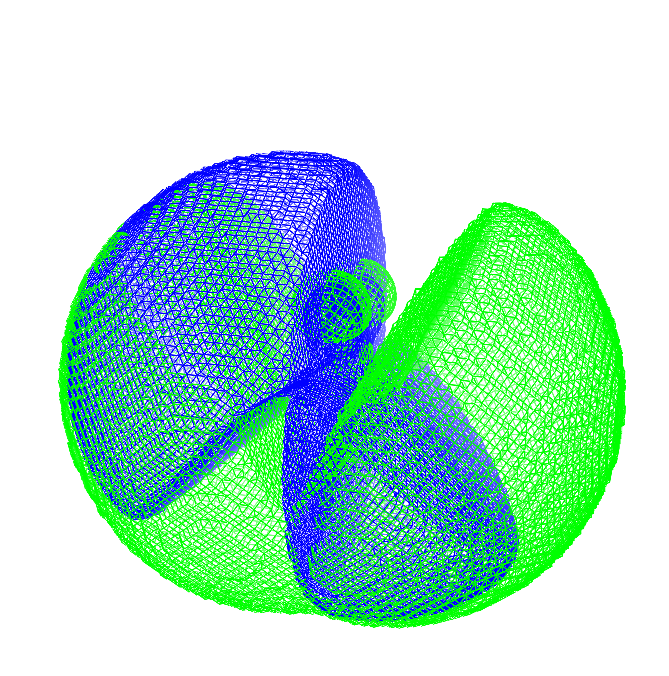}
  }
    \subfigure[R12 and R12$^{+}$ HOMO]{
    \includegraphics[width=0.41\textwidth]{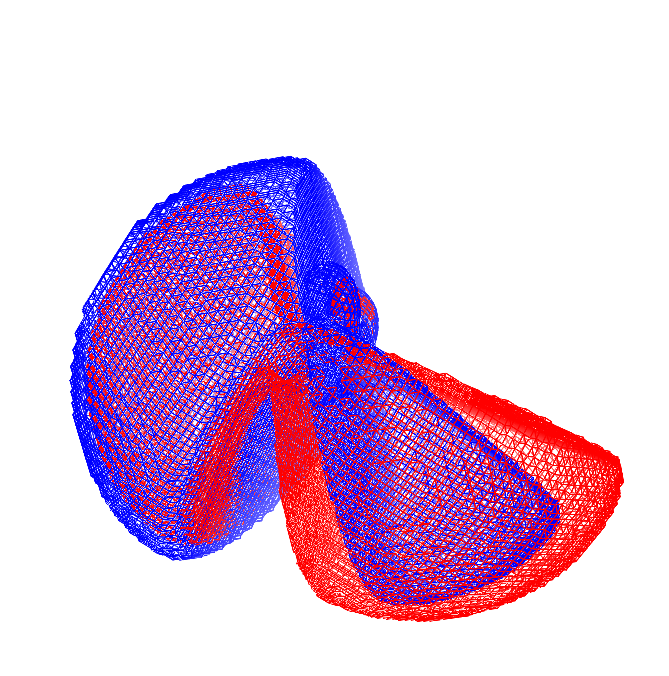}
  }
    \caption{ A comparison of R12$(CCl_2F_2)$ HOMO (in blue) and LUMO (in green) for the neutral molecule (left plot) and the HOMO for neutral R12 molecule (in blue) and charged one R12 ion (in red, right plot) . }
    \label{fig:r12ion}
  \end{center}
\end{figure}

\begin{figure}[]
  \begin{center}
    \subfigure[HFO1234ze HOMO and LUMO]{
    \includegraphics[width=0.41\textwidth]{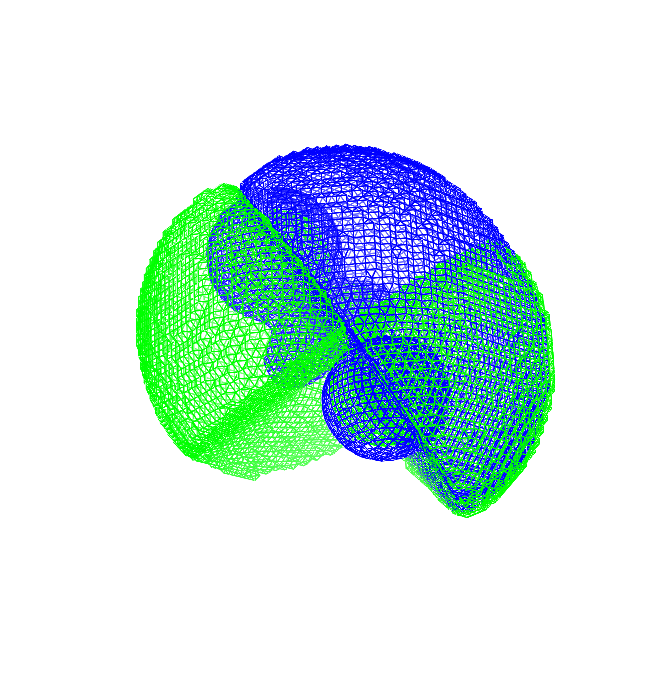}
  }
    \subfigure[HFO1234ze and HFO1234ze$^{+}$ HOMO]{
    \includegraphics[width=0.41\textwidth]{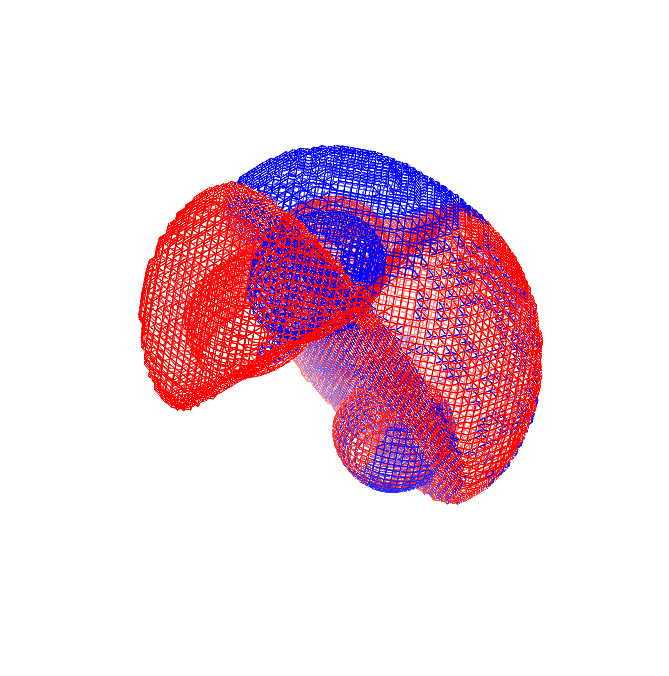}
  }
    \caption{ A comparison of HFO1234ze($CFHCHCF_3$) HOMO (in blue) and LUMO (in green) for the neutral molecule (left plot) and the HOMO for neutral HFO1234ze molecule (in blue) and charged one HFO1234ze ion (in red, right plot) . }
    \label{fig:hfo1234zeion}
  \end{center}
\end{figure}

The Fig.(\ref{fig:r12ion}) compares the HOMO (in blue) and LUMO (in green) of R12 on the left plot and the HOMO for neutral R12 molecule (in blue) and charged one R12 ion (in red) on the right plot. Similar plot for HFO1234ze is shown in Fig.(\ref{fig:hfo1234zeion}). 
As can be seen from the HOMO plot, the molecular orbit between the neutral molecule and charged molecule are similar, especially for the R12, which has less atoms in the molecule. Since the Koopman's theorem will be exact when the molecular orbit for the neutral molecule and charged molecule are the same, we calculated the atom number averaged difference between our two methods, which reflect the difference of the molecular orbitals. The results are shown in Table.(\ref{tab:EA}) 

As the lowest excitation energy can be estimated from the energy gap between the HOMO and LUMO, we also make a comparison of these results to those obtained in the previous section. From the comparison in Table.(\ref{tab:EA}), we see that, although the absolute values of the excitation energy have relatively large differences between the two approaches, the difference of the excitation energies among the molecules considered is only marginal.


\begin{landscape}
\begin{table*}[h!]
  \begin{center}
   \begin{footnotesize}
    \begin{tabular}{|c|c|c|c|c|c|c|c|c|}
      \hline
      Molecule &  GS energy     & GS energy     &  IE1 from    &  HOMO      &   HOMO        &   IEa        &   IEb        &   (IE1-IEm)/N  \\
               &  Neutral [au.] & Charged [au.]     &  GS [eV]     &  HF [au.]  &   DFT [au.]   &   HF [eV]    &  DFT [eV]    &     [eV]       \\
      \hline
      \hline
$CH_4$			&-40.20		&-39.77		&11.78	&-0.51	&-0.52	&13.99	&14.21	&0.46\\
$CF_4$			&-435.65	&-435.18	&12.61	&-0.57	&-0.58	&15.65	&15.65	&0.61\\
R12
   &-1150.28	&-1149.90	&10.24	&-0.39	&-0.39	&10.72	&10.72	&0.10\\
R134a
   &-474.66	&-474.21	&12.40	&-0.38	&-0.38	&10.29	&10.28	&0.26\\
R152a 
   &-276.95	&-276.55	&10.78	&-0.37	&-0.37	&10.15	&10.06	&0.09\\
$CF_3I$	      &-7222.45   &-7222.10	&9.62 	&-0.34	&-0.34	&9.12 	&9.12 	&0.05\\
HFO1234ze 
   &-512.50	&-512.16	&9.34	&-0.29	&-0.29	&7.89	&7.88	&0.16\\
HFO1234yf
   &-512.50	&-512.16	&9.37	&-0.29	&-0.30	&7.84	&8.03	&0.16\\
HFO1233zd 
   &-971.39	&-971.05	&9.23	&-0.25	&-0.25	&6.82	&6.82	&0.27\\
      \hline
    \end{tabular}
  \caption{Results of ionization energy calculation. GS stands for Ground state, IE1 stands for the IE calculated from the GS difference, N stands for the number of atoms in the molecule, and IEm stands for the mean value of IEa, IEb, which are IE calculated with the HF and DFT method accordingly.  }
  \label{tab:IE}
   \end{footnotesize}
  \end{center}
\end{table*}
\end{landscape}

\begin{landscape}
\begin{table*}[h!]
  \begin{center}
   \begin{footnotesize}
    \begin{tabular}{|c|c|c|c|c|c|c|c|c|}
      \hline
  Molecule &  LUMO     & LUMO	       &  HOMO         &  HOMO       &   EA	   &   EA	   &  Lowest Excitation    &  Lowest Excitation \\
           &  HF [au.] & DFT [au.]     &  HF [au.]     &  DFT [au.]  &   HF [eV]   &   DFT [eV]    &  -(HOMO-LUMO)[eV]    &  Spectrum  [eV]       \\
      \hline
      \hline
$CH_4$			&0.16	&0.09	&-0.51	&-0.52	&-4.40	&-2.53	&17.56	&11.17\\
$CF_4$			&0.35	&0.35	&-0.57	&-0.58	&-9.53	&-9.54	&25.18	&17.02\\
R12 
   &0.09	&0.09	&-0.39	&-0.39	&-2.32	&-2.32	&13.04	&10.69\\
R134a 
   &0.16	&0.16	&-0.38	&-0.38	&-4.45	&-4.47	&14.74	&13.11\\
R152a 
   &0.18	&0.18	&-0.37	&-0.37	&-4.93	&-4.96	&15.05	&14.66\\
$CF_3I$     	&0.07	&0.07	&-0.34	&-0.34	&-1.87	&-1.87	&11.00	&8.21\\
HFO1234ze
   &0.05	&0.05	&-0.29	&-0.29	&-1.46	&-1.49	&9.36	&8.21\\
HFO1234yf
   &0.05	&0.06	&-0.29	&-0.30	&-1.40	&-1.58	&9.42	&8.21\\
HFO1233zd
   &0.07	&0.07	&-0.25	&-0.25	&-1.90	&-1.91	&8.72	&6.94\\
      \hline
    \end{tabular}
  \end{footnotesize}
  \caption{Results of electron affinity calculation and the lowest excitation. The lowest excitation energy from the absorption for all molecules except for the $CH_4$, which is directly read out from the calculation, are obtained from the mean value of a Gaussian fit to the lowest peak in the spectrum. }
  \label{tab:EA}
  \end{center}
\end{table*}
\end{landscape}

\section{Estimation of the first Townsend parameter}

Following Ref. \cite{2006ITNS...53.2931D}, we express the dependance of the first Townsend parameter $\alpha$ as a function of the reduced electric field $E/p$ 
\begin{equation}
\dfrac{\alpha}{p} = A e^{-Bp/E}
\end{equation}
where A and B are parameters depending on the gas type and
electric field range, and $p$ the gas pressure.

The reference~\cite{2006ITNS...53.2931D} shows that the first
Townsend coefficient at high reduced electric field depends almost
entirely on the mean free path of the electrons. 
The mean free path, which is defined as 

\begin{equation}
\lambda_{m} = \dfrac{1}{n\sigma},
\end{equation}
where n is the number of atoms per unit volume and $\sigma$ is the
total cross section for electron collision with atoms, can be calculated if the environment of gas molecule is provided. 
Firstly, the n is dependent on the gas pressure and can be put by hand. Secondly, the $\sigma(v)$, which is a function of electron velocity $v$, 
can be calculated from the NIST website~\cite{NIST}.
Then, noting that at high reduced electric field, the first Townsend coefficient can be expressed
as a combination of two components:

\begin{eqnarray}
\alpha_1 = Ap \cdot exp\left(\dfrac{-Bp}{E}\right), E < E_m \\ \nonumber
\alpha_2 = n \sigma, E > E_m
\end{eqnarray}

where $E_m$ is the electric field that drifting an electron over $\lambda_m$ and reach the ionization energy ($I_0$) of the gas. 
Therefore, we can estimate with two free parameters A, B, which need to be determined from experiments. 

For real gases, we cannot take the free path lengths as a constant and the ionization cross section is only a fraction of the total cross section. In that case the estimation needs to be modified and use the following equation:

\begin{equation}
\alpha(r) = Ap \cdot exp\left(\dfrac{-Bp}{E(r)}\right)\left( 1-exp\left( -\dfrac{I_0n\sigma}{eE(r)} \right) \right) + n \sigma_i exp\left( -\dfrac{I_0n\sigma}{eE(r)} \right).
\end{equation}
Numerous measurements of the Townsend coefficients are available for standard gas mixtures, such as those reviewed in \cite{Sharma:1992np}.
\section*{Conclusions}
Currently used F-based gases today used in HEP gas detectors are being phased out by industry and replaced by eco-friendly substitute gases. This study has reported on a general survey of industrially available replacements for HEP gases, discussed their physical properties, materials compatibility and safety issues. Parameters of interest for their use in HEP detectors have been computed following different approaches ranging from parametrizations to quantum chemical calculations: ionization energy, electronegativity, number of primary pairs. Statistical methods to compute amplification parameters of the ionization shower production such as the Townsend coefficients were investigated and preliminary results reported. Promising candidates with lower GWP are identified for further studies.
\section*{Acknowledgements}
This work was funded by Istituto Nazionale di Fisica Nucleare and Ministero della Istruzione, Universit\`a e Ricerca of Italy, the Center for Research in Nuclear Physics, the Peking University of China.
Discussions with and suggestions from Marcello Abbrescia (Universit\`a di Bari and INFN) are gratefully acknowledged.
\newpage
\section{Appendix - Datasheets}

\begin{landscape}
\begin{table*}[h!]
  \begin{center}
   \begin{footnotesize}
    \begin{tabular}{|l|}
      \hline
 {\bf R125} \\
    National Refrigerants.   http:$/$$/$www.refrigerants.com$/$msds$/$r125.pdf \\
      Dupont  http:$/$$/$www2.dupont.com$/$inclusive-innovations$/$en-us$/$sites$/$default$/$files$/ $Pentafluoroethane\%20\%28R125\%29\%20Product\%20Safety\%20Summary.pdf \\
      http:$/$$/$www.raprec.com$/$download$/$resources$/$msds$/$R-125\%20MSDS.pdf \\
    BOC gases https:$/$$/$www.boconline.co.uk$/$internet.lg.lg.gbr$/$en$/$images$/$r125410\_55984.pdf \\
    {\bf R23} \\
         Matheson trigas 
      https:$/$$/$www.mathesongas.com$/$pdfs$/$msds$/$MAT09970.pdf\\
        AIRGAS   https:$/$$/$www.airgas.com$/$msds$/$001078.pdf \\
       PRAXAIR  http:$/$$/$www.praxair.com$/$~$/$media$/$North\%20America$/$US$/$Documents$/$SDS$/$Halocarbon\%2023\%20CHF3\%20Safety\%20Data\%20Sheet\%20SDS\%20P4668.ashx\\
     Arkema           http:$/$$/$www.lskair.com$/$MSDS$/$R23\%20MSDS.pdf \\
     National Refrigerants.  http:$/$$/$www.refrigerants.com$/$msds$/$r23.pdf\\
       BOC gases  https:$/$$/$www.boconline.co.uk$/$internet.lg.lg.gbr$/$en$/$images$/$r23410\_64635.pdf \\
        Airliquid  http:$/$$/$www.first.ethz.ch$/$infrastructure$/$Chemicals$/$Gases$/$MSDS\_TriFluoroMethane.pdf \\
     {\bf R143}  \\
         Dupont  http:$/$$/$www.raprec.com$/$download$/$resources$/$msds$/$R-143a\%20MSDS.pdf \\
         Air Liquide   http:$/$$/$www.msds-al.co.uk$/$assets$/$file\_assets$/$SDS\_118-CLP-TRIFLUOROETHANE\_R143a.pdf \\
            Air Liquide Gas Encyclopedia http:$/$$/$encyclopedia.airliquide.com$/$encyclopedia.asp?GasID=114 \\
            Rivoira  http:$/$$/$www.rivoiragas.it$/$wp-content$/$uploads$/$files$/$r407c\%202009\%20en.pdf \\
     {\bf R14} \\
           AIRGAS      https:$/$$/$www.airgas.com$/$msds$/$002847.pdf \\
           PRAXAIR http:$/$$/$www.praxair.com$/$~$/$media$/$North\%20America$/$US$/$Documents$/$SDS$/ $Halocarbon \%2014\%20CF4\%20Safety\%20Data\%20Sheet\%20SDS\%20P4665.ashx\\
       Matheson trigas https:$/$$/$www.mathesongas.com$/$pdfs$/$msds$/$MAT02100.pdf \\
       Lynde  http:$/$$/$www.linde-gas.it$/$internet.lg.lg.ita$/$it$/$images$/$1062\_R14\_CF4335\_68692.pdf \\
         Air Liquide    http:$/$$/$www.msdshazcom.com$/$MSDS$/$A$/$Air\%20Liquide$/$116\_AL\_EN\_Tetrafluoromethane\%20\%28R14\%29.pdf \\
        BOC gases http:$/$$/$www.google.ch$/$url?sa=t\&rct=j\&q=\&esrc=s\&source=web\&cd=2\&ved=0CCQQFjAB\&url=http\%3A\%2F\%2Fwww.chemicalsfinder.com\%2Fdownload.php\%3Ffile \\
        \%3DTVNEUy8xNTQ5M19CT0MtR2FzZXMtVUtfSGFsb2NhcmJvbi0xNF92Mi5wZGY\%3D\&ei=shoQVPTSHInTaPXdgfAJ\&usg=AFQjCNHT1X1rbxqJpstbO8oM9G1M4wSCWg\&sig2=kzAZpbdDJ6406R860FyCew \\
      Spectra Gases Material safety data sheet http:$/$$/$www.spectragases.com.cn$/$pdfs$/$MSDS$/$pure\%20gases$/$MSDS\_Tetrafluoromethane-1028\_121305.pdf \\
   {\bf R7146} \\
    Airgas Material safety data sheet
https:$/$$/$www.airgas.com$/$msds$/$001048.pdf \\
     PRAXAIR safety data sheet http:$/$$/$www.praxair.com$/$~$/$media$/$North\%20America$/$US$/$Documents$/$SDS$/$Sulfur\%20Hexafluoride\%20SF6\%20Safety\%20Data\%20Sheet\%20SDS\%20P4657.ashx \\
    Air Liquide material data sheet http:$/$$/$mfc.engr.arizona.edu$/$safety$/$MSDS\%20FOLDER$/$SF6\%20-\%20MSDS\%20Air\%20Liquide.pdf \\
       Concorde material data sheet http:$/$$/$www.concordegas.com$/$Images-\%281\%29$/$pdf$/$SF6-MSDS-English.aspx \\
     AFROX material data sheet http:$/$$/$www.afrox.co.za$/$internet.global.corp.zaf$/$en$/$images$/$Sulphur\_Hexafluoride266\_27767.pdf \\
  {\bf R407c} \\
   National Refrigerants.  Material safety data sheet        http:$/$$/$www.refrigerants.com$/$msds$/$r407c.pdf \\
     Rivoira Material safety data sheet http:$/$$/$www.rivoiragas.it$/$wp-content$/$uploads$/$files$/$r407c\%202009\%20en.pdf \\
      Honeywell Material safety data sheet http:$/$$/$msds-resource.honeywell.com$/$ehswww$/$hon$/$result$/$
result\_single.jsp?P\_LANGU=E\&P\_SYS=1\&C001= \\
 MSDS\&C997=C100\%3BESDS\_US\%2BC102\%3BUS\%2B1000\&C100=*\&C101=*\&C102=*\&C005=000000009894\&C008=\&C006=HON\&C013=+ \\
     Harp Material safety data sheet http:$/$$/$www.harpintl.com$/$downloads$/$pdf$/$msds$/$harp-r407c-sds-clp.pdf \\
       BOC gases Material safety data sheet https:$/$$/$www.boconline.co.uk$/$internet.lg.lg.gbr$/$en$/$images$/$r407c410\_55995.pdf \\
   {\bf R142b} \\
    AIRGAS Material safety data sheet https:$/$$/$www.airgas.com$/$msds$/$007729.pdf \\
      National Refrigerants.  Material safety data sheet          http:$/$$/$www.refrigerants.com$/$msds$/$nri-142b.pdf \\
     BOC gases Material safety data sheet             https:$/$$/$www.boconline.co.uk$/$internet.lg.lg.gbr$/$en$/$images$/$r142b410\_55987.pdf \\
     Air Liquide material data sheet            http:$/$$/$www.msds-al.co.uk$/$assets$/$file\_assets$/$SDS\_025-CLP- CHLORODIFLUOROETHANE.pdf\\
     Solvay Chemical material data sheet 
                 http:$/$$/$www.solvaychemicals.us$/$SiteCollectionDocuments$/$sds$/$P19446-USA\_CN\_EN.pdf\\  
                 \hline
    \end{tabular}
  \caption{Gas data sheets (1). }
  \label{tab:GDS1}
   \end{footnotesize}
  \end{center}
\end{table*}
\end{landscape}

\begin{landscape}
\begin{table*}[h!]
  \begin{center}
   \begin{footnotesize}
    \begin{tabular}{|l|}
      \hline
  {\bf R134a} \\
Honeywell Material safety data sheet https:$/$$/$www.conncoll.edu$/$media$/$website-media$/$offices$/$ehs$/$
envhealthdocs$/$Genetron\_R-134a.pdf \\
National Refrigerants.  Material safety data sheet http:$/$$/$www.refrigerants.com$/$msds$/$r134a.pdf \\
   {\bf R124} \\
AIRGAS Material safety data sheet https:$/$$/$www.airgas.com$/$msds$/$001135.pdf \\
Arkema Material Safety Data Sheet http:$/$$/$www.hudsontech.com$/$wp-content$/$themes$/$hudson$/$pdfs$/$
msds$/$R-124$/$ARKEMA\_R-124.pdf \\
Honeywell Material safety data sheet http:$/$$/$www.honeywell.com$/$sites$/$docs$/$doc19194b8-fb3eb4d751-3e3e4447ab3472a0c2a5e5fdc1e6517d.pdf \\
BOC gases Material safety data sheet
https:$/$$/$www.boconline.co.uk$/$internet.lg.lg.gbr$/$en$/$images$/$r124410\_55983.pdf \\
Actrol Material safety data sheet
http:$/$$/$webcache.googleusercontent.com$/$search?q=cache:KUwNl9lnEIUJ:https:$/$$/$go.lupinsys.
com$/$actrol$/$harms$/$public$/$materials$/$85cd6647d2c675a5fc02bf5367e562d8-published\\
$/$attachments\_api$/$997a373e34311b7c499864235208d0ef$/$search\_api$/$R124\_Refrigerant-MSDS.pdf+\&cd=5\&hl=it\&ct=clnk\&gl=ch\&client= firefox-a \\
Advanced Specialty Gases Material safety data sheet
http:$/$$/$www.advancedspecialtygases.com$/$pdf$/$R-124\_MSDS.pdf \\
Honeywell Material safety data sheet
http:$/$$/$www.3eonline.com$/$ImageServer$/$NewPdf$/$9c88917e-07a6-4217-ac6e-07efa24eabb0$/$9c88917e-07a6-4217-ac6e-07efa24eabb0.pdf \\
   {\bf R123} \\
Global Refrigerant Material safety data sheet
http:$/$$/$www.globalrefrigerants.com.sg$/$docs\%5Cglobal\_r123\_msds.pdf \\
National Refrigerants.  Material safety data sheet http:$/$$/$www.refrigerants.com$/$msds$/$nri-r123.pdf \\
BOC gases Material safety data sheet
https:$/$$/$www.boconline.co.uk$/$internet.lg.lg.gbr$/$en$/$images$/$r123410\_55982.pdf \\
AFROX material data sheet
http:$/$$/$www.afrox.co.za$/$internet.global.corp.zaf$/$en$/$images$/$R123266\_27717.pdf \\
AIRGAS Material safety data sheet
https:$/$$/$www.inrf.uci.edu$/$wordpress$/$wp-content$/$uploads$/$trifluoroethaneHalocarbon-r-123.pdf \\
Air Liquide material data sheet 
http:$/$$/$docs.airliquide.com.au$/$MSDSCalgaz$/$50106.pdf \\
Refrigerant inc. material data sheet
http:$/$$/$www.refrigerantsinc.com$/$images$/$R123\_MSDS.pdf \\
Honeywell Material safety data sheet
http:$/$$/$msds-resource.honeywell.com$/$ehswww$/$hon$/$result$/$result\_single.jsp?P\_LANGU\\
=E\&P\_SYS=1\&C001=MSDS\&C997=C100\%3BESDS\_US\%2BC102\%3BUS\%2B1000\&C100=*\&C101=*\&C102=*\&C005=000000009885\&C008=\&C006=HON\&C013=+ \\
Honeywell Material safety data sheet
http:$/$$/$www.3eonline.com$/$ImageServer$/$NewPdf$/$5cb53264-dd88-47f3-b77d-7fbff54287d7$/$5cb53264-dd88-47f3-b77d-7fbff54287d7.pdf \\
  {\bf R290} \\
National Refrigerants.  Material safety data sheet
http:$/$$/$www.refrigerants.com$/$MSDS$/$nri-R290.pdf \\
Kaltis Material safety data sheet
http:$/$$/$www.kaltis.com$/$PDF$/$R290\%20\_Propane\_\%20MSDS.pdf \\
Global Refrigerant Material safety data sheet
http:$/$$/$www.globalrefrigerants.com.sg$/$docs$/$global\_r290\_msds.pdf \\
Lynde Material safety data sheet
http:$/$$/$www.lindecanada.com$/$internet.lg.lg.can$/$en$/$images$/$Propane\_EN135\_104332.pdf \\
AIRGAS Material safety data sheet
https:$/$$/$www.airgas.com$/$msds$/$001135.pdf \\
AmeriGas Propane, L.P. Material safety data sheet
http:$/$$/$www.amerigas.com$/$pdfs$/$AmeriGas-Propane-MSDS.pdf \\
BOC gases Material safety data sheet
https:$/$$/$www.boconline.co.uk$/$internet.lg.lg.gbr$/$en$/$images$/$r124410\_55983.pdf \\
Advanced Gas Technologies Material safety data sheet
http:$/$$/$www.fieldenvironmental.com$/$assets$/$files$/$MSDS\%20Sheets$/$MSDS\%20Sheets\%202012$/$4\%20CarbonDioxide.pdf \\
AFROX material data sheet
http:$/$$/$www.afrox.co.za$/$internet.global.corp.zaf$/$en$/$images$/$R409A266\_27725.pdf \\
Honeywell Material safety data sheet
http:$/$$/$www.honeywell.com$/$sites$/$docs$/$doc19194b8-fb3eb4d751-3e3e4447ab3472a0c2a5e5fdc1e6517d.pdf \\
Energas Material safety data sheet
http:$/$$/$energas.co.uk$/$pdfs$/$013-Migweld.pdf \\
Calor Material safety data sheet
https:$/$$/$www.calor.co.uk$/$media$/$wysiwyg$/$PDF$/$propane\_safety\_data\_sheet.pdf \\
Unitor Material safety data sheet
http:$/$$/$www.wilhelmsen.com$/$services$/$maritime$/$companies$/$buss$/$DocLit$/$MaterialSafety$/$Documents$/$MSDS$/$Refrigeration$/$REFRIGERANT\_R\_290\_Italian.pdf \\
  {\bf R125} \\
National Refrigerants.  Material safety data sheet
http:$/$$/$www.refrigerants.com$/$msds$/$r125.pdf \\
Dupont Material  data sheet
http:$/$$/$www2.dupont.com$/$inclusive-innovations$/$en-us$/$sites$/$default$/$files$/$Pentafluoroethane\%20\%28R125\%29\%20Product\%20Safety\%20Summary.pdf \\
BOC gases Material safety data sheet
https:$/$$/$www.boconline.co.uk$/$internet.lg.lg.gbr$/$en$/$images$/$r125410\_55984.pdf \\
AFROX material data sheet
http:$/$$/$www.afrox.co.za$/$internet.global.corp.zaf$/$en$/$images$/$NAF\_S125266\_27697.pdf \\
                 \hline
    \end{tabular}
  \caption{Gas data sheets (2). }
  \label{tab:GDS2}
   \end{footnotesize}
  \end{center}
\end{table*}
\end{landscape}

\begin{landscape}
\begin{table*}[h!]
  \begin{center}
   \begin{footnotesize}
    \begin{tabular}{|l|}
      \hline
   {\bf R218} \\
AIRGAS 
http:$/$$/$webcache.googleusercontent.com$/$search?q=cache:TdB2lKIW34AJ:www.airgasrefrigerants.com\\$/$admin$/$includes$/$doc\_view.php\%3FID\%3D242+\&cd=8\&hl=it\&ct=clnk\&gl=ch\&client=firefox-a \\
Spectra Gases 
http:$/$$/$www.spectragases.com.cn$/$pdfs$/$MSDS$/$pure\%20gases$/$MSDS\_Octafluoropropane1059\_020507.pdf \\
BOC gases 
https:$/$$/$www.boconline.co.uk$/$internet.lg.lg.gbr$/$en$/$images$/$perfluoropropane410\_64742.pdf \\
Tygris 
http:$/$$/$www.tygrisindustrial.com$/$product$/$static$/$assets$/$downloads$/$coshh$/$R218\_5060253470215.pdf \\
Lynde 
http:$/$$/$msds.lindeus.com$/$files$/$msds$/$WPS\_LIND\_022\_NA\_MSDS\_FINAL\_REV\_8\_31\_10.pdf \\
    {\bf R318} \\
Lynde 
http:$/$$/$msds.lindeus.com$/$files$/$msds$/$WPS\_LIND\_021\_NA\_MSDS\_FINAL\_REV\_8\_31\_10.pdf \\
Matheson tri-gas 
http:$/$$/$www.nfc.umn.edu$/$assets$/$pdf$/$msds$/$octafluorocyclobutane.pdf \\
Spectra Gases 
http:$/$$/$www.spectragases.com.cn$/$pdfs$/$MSDS$/$pure\%20gases$/$MSDS\_Octafluorocyclobutane-1057\_121305.pdf \\
Carboline 
http:$/$$/$www.google.ch$/$url?sa=t\&rct=j\&q=\&esrc=s\&source=web\&cd=4\&ved=0CDYQFjAD\&url=http\%3A\%2F\%2Fmsds.carboline.com \\
\%2Fservlet\%2FFeedFile\%2F1\%2Fprod\%2FR318\%2FBB3C031D1577B6D885257B830054E154\%2FR318B1NL\_2\_USANSI.pdf \\
\&ei=9lwRVO35BJPnaKW7geAO\&usg=AFQjCNGHlgVYKj9--VNkldq-WxyPfMJN8g\&sig2=wQ871i2BYRipK5OffyzBMw \\
  {\bf R600a} \\
Harp International
http:$/$$/$www.harpintl.com$/$downloads\%5Cpdf\%5Cmsds\%5CHARP-R600a-CLP.pdf \\
National Refrigerants.  
http:$/$$/$www.refrigerants.com$/$MSDS$/$nri-R600a.pdf \\
Global refrigerants 
http:$/$$/$globalrefrigerants.com.sg$/$docs$/$global\_r600a\_msds.pdf \\
Kaltis 
http:$/$$/$www.kaltis.com$/$PDF$/$R600a\%20\_Isobutane\_\%20MSDS.pdf \\
Frigostar Refrigerants 
http:$/$$/$www.soos.hu$/$WEBSET\_DOWNLOADS$/$603$/$R600a\_-\_MSDS\_EN.pdf \\
PRAXAIR 
http:$/$$/$www.praxair.com$/$~$/$media$/$North\%20America$/$US$/$Documents$/$SDS$/$Isobutane\%20C4H10\%20Safety\%20Data\%20Sheet\%20SDS\%20P4613.ashx \\
A-Gas UK Limited 
http:$/$$/$www.climatecenter.co.uk$/$wcsstore7.00.00.749$/$ExtendedSitesCatalogAssetStore$/$images$/$products$/$AssetPush$/$DTP\_AssetPushHighRes$/$std.lang.all$/$\_h$/$\&s$/$R600a\_Isobutane\_H\&S.pdf \\
 {\bf HFO1234ze} \\
Honeywell 
http:$/$$/$msds-resource.honeywell.com$/$ehswww$/$hon$/$result$/$result\_single.jsp?P\_LANGU=E\&P\_SYS=1\&C001=MSDS\&C997=C100\%3BE\%2BC \\
101\%3BSDS\_US\%2BC102\%3BUS\%2B1000\&C100=*\&C101=*\&C102=*\&C005=000000014785\%20\%20\&C008=\&C006=HON\&C013=+ \\
Harp International Material safety data sheet
http:$/$$/$www.harpintl.com$/$downloads$/$pdf$/$msds$/$Solstice-1234ze-SDS-CLP.pdf \\
 {\bf R13I1} \\
Matheson tri-gas safety data sheet
https:$/$$/$www.mathesongas.com$/$pdfs$/$msds$/$00229444.pdf \\
IODEAL Brand safety data sheet
http:$/$$/$coupp-docdb.fnal.gov$/$cgi-bin$/$RetrieveFile?docid=163;filename=2.3\%20CF3I\%20MSDS.pdf;version=2 \\
Lookchem safety data sheet
http:$/$$/$www.lookchem.com$/$msds$/$2011-06\%2F2\%2F171441\%282314-97-8\%29.pdf \\
National Industrial Chemicals Notification and Assessment Scheme (NICNAS)
http:$/$$/$www.nicnas.gov.au$/$\_\_data$/$assets$/$pdf\_file$/$0011$/$9101$/$NA334FR.PDF \\
AlfaAesar safety data sheet
http:$/$$/$www.alfa.com$/$content$/$msds$/$british$/$41479.pdf \\
http:$/$$/$www.alfa.com$/$content$/$msds$/$british$/$39665.pdf \\
Strem Chemicals, Inc safety data sheet
http:$/$$/$www.strem.com$/$catalog$/$msds$/$09-7680 \\
Pfaltz - Bauer safety data sheet
https:$/$$/$www.pfaltzandbauer.com$/$MSDS$/$T23585\%20\%20SDS\%20\%20051613.PDF \\
 {\bf HFO1233zd} \\
Honeywell safety data sheet
http:$/$$/$www.ozoneprogram.ru$/$upload$/$files$/$2$/$2013\_honeywell.pdf \\
{\bf HFO1234yf} \\
Honeywell 
http:$/$$/$msds-resource.honeywell.com$/$ehswww$/$hon$/$result$/$result\_single.jsp?P\_LANGU=E\&P\_SYS=1\&C001=MSDS\&C997 \\
=C100;E\%2BC101;SDS\_US\%2BC102;US\%2B1000\&C100=*\&C101=*\&C102=*\&C005=000000011078\&C008\&C006=HON\&C013 \\
http:$/$$/$www.idsrefrigeration.co.uk$/$docs$/$Solstice\%20yf\%20MSDS\%20version\%203.2.pdf \\
Harp International Material safety data sheet
http:$/$$/$www.harpintl.com$/$downloads$/$pdf$/$msds$/$harp-hfo-1234yf-clp.pdf \\
BOC gases Material safety data sheet
https:$/$$/$www.boconline.co.uk$/$internet.lg.lg.gbr$/$en$/$images$/$tetrafluoropropene-r1234yf-sg-155410\_90007.pdf \\
National Refrigerants  Material safety data sheet
http:$/$$/$www.nationalref.com$/$pdf$/$19\%20SDSR1234yf.pdf \\
Gerling Holz+co Material safety data sheet
http:$/$$/$www.ghc.de$/$media$/$en$/$downloads$/$msds$/$0070.pdf \\
Friedrichs K\"a\%ltemittel Gmbh
http:$/$$/$www.friedrichs-kaeltemittel.de$/$download$/$deutsch$/$kaeltemittel$/$r1234yf$/$ENG\%20-\%20R1234yf-2013-08-09\%20-\%20Friedrichs\%20Kaeltemittel.pdf \\
                 \hline
    \end{tabular}
  \caption{Gas data sheets (3). }
  \label{tab:GDS3}
   \end{footnotesize}
  \end{center}
\end{table*}
\end{landscape}

\begin{landscape}
\begin{table*}[h!]
  \begin{center}
   \begin{footnotesize}
    \begin{tabular}{|l|}
      \hline
{\bf R116} \\
Matheson tri-gas safety data sheet
https:$/$$/$www.mathesongas.com$/$pdfs$/$msds$/$MAT10860.pdf \\
PRAXAIR safety data sheet
http:$/$$/$www.praxair.com$/$~$/$media$/$North\%20America$/$US$/$Documents$/$SDS$/$Halocarbon\%20116\%20C2F6\%20Safety\%20Data\%20Sheet\%20SDS\%20P4670.ashx \\
HYNOTE GAS safety data sheet
http:$/$$/$www.uacj.mx$/$IIT$/$CICTA$/$Documents$/$Gases$/$Hexafluoroethane.pdf \\
Spectra Gases Material safety data sheet
http:$/$$/$www.spectragases.com.cn$/$pdfs$/$MSDS$/$pure\%20gases$/$MSDS\_Hexafluoroethane-1058\_092605.pdf \\
BOC gases Material safety data sheet
https:$/$$/$www.boconline.co.uk$/$internet.lg.lg.gbr$/$en$/$images$/$sg\_064\_hexafluoroethane\_r116410\_64639.pdf \\
National Refrigerants  Material safety data sheet
http:$/$$/$www.refrigerants.com$/$msds$/$r116.pdf \\
AIRGAS Material safety data sheet
https:$/$$/$www.airgas.com$/$msds$/$001053.pdf \\
Dupont Material  data sheet
http:$/$$/$www.raprec.com$/$download$/$resources$/$msds$/$R-116\%20MSDS.pdf \\
Air Products and Chemicals,Inc Material  data sheet
http:$/$$/$cleanroom.ien.gatech.edu$/$media$/$uploads$/$msds$/$350.pdf \\
PRAXAIR Material safety data sheet
http:$/$$/$www.praxair.com$/$~$/$media$/$North\%20America$/$US$/$Documents$/$SDS$/$Halocarbon\%20116\%20C2F6\%20Safety\%20Data\%20Sheet\%20SDS\%20P4670.ashx \\
SPECIALTY CHEMICAL PRODUCT Material  data sheet
http:$/$$/$www.chemadvisor.com$/$Matheson$/$database$/$msds$/$00232361000800003.PDF \\
{\bf R32} \\
National Refrigerants  Material safety data sheet
http:$/$$/$www.refrigerants.com$/$msds$/$nri-r32.pdf \\
BOC gases Material safety data sheet
https:$/$$/$www.boconline.co.uk$/$internet.lg.lg.gbr$/$en$/$images$/$sg\_152-r32-difluoromethane-v1.51410\_39651.pdf \\
AIRGAS Material safety data sheet
https:$/$$/$www.airgas.com$/$msds$/$001054.pdf \\
Spectra Gases Material safety data sheet
http:$/$$/$www.spectragases.com.cn$/$pdfs$/$MSDS$/$pure\%20gases$/$MSDS\_Difluoromethane-1029\_121305.pdf \\
Matheson tri-gas safety data sheet
https:$/$$/$www.mathesongas.com$/$pdfs$/$msds$/$MAT14937.pdf \\
Harp International Material safety data sheet
http:$/$$/$www.harpintl.com$/$downloads$/$pdf$/$msds$/$harp-r410a-sds-clp.pdf \\
Dupont Material  data sheet
http:$/$$/$www2.dupont.com$/$inclusive-innovations$/$en-us$/$sites$/$default$/$files$/$Difluoromethane\%20\%28R32\%29\%20Product\%20Safety\%20Summary.pdf \\
{\bf R115} \\
National Refrigerants  Material safety data sheet
http:$/$$/$www.refrigerants.com$/$msds$/$nri-r115.pdf \\
AIRGAS Material safety data sheet
https:$/$$/$www.airgas.com$/$msds$/$009900.pdf \\
Air Liquide material data sheet 
http:$/$$/$www.msds-al.co.uk$/$assets$/$file\_assets$/$SDS\_030-CLP-CHLOROPENTAFLUOROETHANE.pdf \\
Dupont Material  data sheet
http:$/$$/$www.raprec.com$/$download$/$resources$/$msds$/$R-115\%20MSDS.pdf \\
Zep Inc. Material  data sheet
http:$/$$/$www.zepprofessional.com$/$msds$/$eng$/$R115\_ENG\_USA.pdf \\
PRAXAIR Material safety data sheet
http:$/$$/$www.praxair.ca$/$~$/$media$/$North\%20America$/$Canada$/$Documents\%20en$/$Safety\%20Data\%20Sheets\%20en$/$Halocarbon\%20115\%20SDS\%20E4669.ashx \\
Apolllo Scientific Limited Material safety data sheet
http:$/$$/$www.apolloscientific.co.uk$/$downloads$/$msds$/$PC2020\_msds.pdf \\
Matheson tri-gas safety data sheet
https:$/$$/$www.mathesongas.com$/$pdfs$/$msds$/$MAT04810.pdf \\
                 \hline
    \end{tabular}
  \caption{Gas data sheets (5). }
  \label{tab:GDS5}
   \end{footnotesize}
  \end{center}
\end{table*}
\end{landscape}

\begin{landscape}
\begin{table*}[h!]
  \begin{center}
   \begin{footnotesize}
    \begin{tabular}{|l|}
      \hline
{\bf R22 } \\
National Refrigerants  Material safety data sheet
http:$/$$/$www.refrigerants.com$/$msds$/$r22.pdf \\
BOC gases Material safety data sheet
https:$/$$/$www.boconline.co.uk$/$internet.lg.lg.gbr$/$en$/$images$/$sg-027-r22-chlorodifluoromethane-v1.3410\_39623.pdf\\
Harp International Material safety data sheet
http:$/$$/$www.harpintl.com$/$downloads$/$pdf$/$msds$/$HARP-R22-SDS-CLP.pdf \\
AIRGAS Material safety data sheet
https:$/$$/$www.airgas.com$/$msds$/$001016.pdf \\
Remtec Material safety data sheet
http:$/$$/$www.remtec.net$/$docs$/$msds-r-22-remtec.pdf \\
AFROX material data sheet
http:$/$$/$www.afrox.co.za$/$internet.global.corp.zaf$/$en$/$images$/$R22266\_27716.pdf \\
Gasco material data sheet
http:$/$$/$site.jjstech.com$/$pdf$/$Gasco$/$MSDS-Freon-R22-in-Air.pdf \\
Dupont Material  data sheet
http:$/$$/$www.pchetz.com$/$\_Uploads$/$dbsAttachedFiles$/$freon22msds.pdf \\
SRF LIMITED Material  data sheet
http:$/$$/$www.siggases.com$/$images$/$products$/$MSDS-\%20FLORON\%20R22.pdf \\
Lynde Material  data sheet
http:$/$$/$msds.lindeus.com$/$files$/$msds$/$WPS\_LIND\_011\_CHLORODIFLUOROMETHANE\_\%28Halocarbon\_R22\%29\_NA\_MSDS\_FINAL\_REV\_8\_25\_10.pdf \\
Unitor Material  data sheet
http:$/$$/$www.wilhelmsen.com$/$services$/$maritime$/$companies$/$buss$/$DocLit$/$MaterialSafety$/$Documents$/$MSDS$/$Refrigeration$/$UNICOOL\_R\_22\_English.pdf \\
A-GAS Material  data sheet
http:$/$$/$www.airefrig.com.au$/$file$/$msds$/$Agas\_R22\_MSDS\_100708.pdf \\
Technical Chemical Company   Material  data sheet
http:$/$$/$www.technicalchemical.com$/$msds$/$6230.pdf \\
Praxair Material  data sheet
http:$/$$/$www.praxair.com$/$~$/$media$/$North\%20America$/$US$/$Documents$/$SDS$/$Halocarbon\%2022\%20CHCIF2\%20Safety\%20Data\%20Sheet\%20SDS\%20P4667.ashx \\
Refron Material  data sheet
http:$/$$/$www.acsrefrigerant.com$/$pdf$/$REFRON\_MSDS\_R-22.pdf \\
Honeywell Material  data sheet
http:$/$$/$www.honeywell.com$/$sites$/$docs$/$doc19194b8-fb3eb861e4-3e3e4447ab3472a0c2a5e5fdc1e6517d.pdf \\
Patton LTD Material  data sheet
http:$/$$/$www.pattonnz.com$/$pdf$/$MSDS$/$MSDS\_2013$/$Refrigerants$/$Refrigerant\%20R22\%20-\%20MSDS.pdf \\
Coregas PTY LTD Material  data sheet
http:$/$$/$www.bjhowes.com.au$/$R22\%20-\%20CHLORODIFLUOROMETHANE\%20REFRIGERANT\%20GAS\%5B1\%5D.pdf \\
{\bf R152} \\
Dupont Material  data sheet
http:$/$$/$www.raprec.com$/$download$/$resources$/$msds$/$R-152a\%20MSDS.pdf \\
BOC gases Material safety data sheet
https:$/$$/$www.boconline.co.uk$/$internet.lg.lg.gbr$/$en$/$images$/$sg-045-r152a-11-difluoroethane-v1.3410\_39625.pdf \\
National Refrigerants  Material safety data sheet
http:$/$$/$www.refrigerants.com$/$msds$/$r152a.pdf \\
Electron Microscopy Sciences Material safety data sheet
http:$/$$/$www.emsdiasum.com$/$microscopy$/$technical$/$msds$/$70837.pdf \\
Matheson tri-gas safety data sheet
https:$/$$/$www.mathesongas.com$/$pdfs$/$msds$/$MAT26280.pdf \\
Lynde Material safety data sheet
http:$/$$/$www.abellolinde.es$/$internet.lg.lg.esp$/$en$/$images$/$Gas\%20Refrigerante\%20R152a\_FDS\_7071\_01\_00FDS\_SG\_045\_01\_00302\_89373.pdf \\
Air liquide  Material safety data sheet
http:$/$$/$www.msds-al.co.uk$/$assets$/$file\_assets$/$SDS\_045-CLP-DIFLUOROETHANE.pdf \\
AIRGAS Material safety data sheet
https:$/$$/$www.airgas.com$/$msds$/$001029.pdf \\
SIDS Initial Assessment Report For SIAM 22 Paris, France, 18-21 April 2006 
http:$/$$/$www.chem.unep.ch$/$irptc$/$sids$/$OECDSIDS$/$75376.pdf \\
\hline
   \end{tabular}
  \caption{Gas data sheets (6). }
  \label{tab:GDS6}
   \end{footnotesize}
  \end{center}
\end{table*}
\end{landscape}

\newpage
\end{document}